\documentclass[11pt]{article}
\usepackage{graphicx}
\usepackage{amssymb}
\usepackage{amscd}
\usepackage{mathrsfs}
\usepackage{longtable,lscape}
\usepackage{amsthm}
\usepackage{amsfonts}
\usepackage{amsmath}
\usepackage{bbm}
\usepackage{float}
\usepackage{url}
\usepackage{hyperref}
\usepackage[margin=0.5cm,font=footnotesize]{caption}
\usepackage{lineno}
\usepackage[round]{natbib}
\usepackage{appendix}
\usepackage{subfig}

\begin{document}
\title{\bf Turing Video-based Cognitive to Handle 
\\ Tests Entangled Concepts}
\author{Diederik Aerts\footnote{Center Leo Apostel for Interdisciplinary Studies, 
        Vrije Universiteit Brussel (VUB), Pleinlaan 2,
         1050 Brussels, Belgium; email address: diraerts@vub.be} 
$\,$ Roberto Leporini\footnote{Department of Economics, University of Bergamo, via dei Caniana 2, 24127 Bergamo, Italy; email address: roberto.leporini@unibg.it} 
        $\,$ and $\,$  Sandro Sozzo\footnote{Department of Humanities and Cultural Heritage (DIUM) and Centre CQSCS, University of Udine, Vicolo Florio 2/b, 33100 Udine, Italy; email address: sandro.sozzo@uniud.it}              }

\date{}
\maketitle
\begin{abstract}
\noindent 
We have proved in both human-based and computer-based tests that natural concepts generally `entangle' when they combine to form complex sentences, violating the rules of classical compositional semantics. In this article, we present the results of an innovative video-based cognitive test on a specific conceptual combination, which significantly violates the Clauser--Horne--Shimony--Holt version of Bell's inequalities (`CHSH inequality'). We also show that collected data can be faithfully modelled within a quantum-theoretic framework elaborated by ourselves and a `strong form of entanglement' occurs between the component concepts. While the video-based test confirms previous empirical results on entanglement in human cognition, our empirical approach surpasses language barriers and eliminates the need for prior knowledge, enabling universal accessibility. Finally, this transformative methodology allows one to unravel the underlying connections that drive our perception of reality. As a matter of fact, we provide a novel explanation for the appearance of entanglement in both physics and cognitive realms.
\end{abstract}

\section{Introduction\label{intro}}
Completely Automated Public Turing test to tell Computers and Humans Apart (CAPTCHA\footnote{The term ``CAPTCHA'' was coined in 2000 by Luis von Ahn, Manuel Blum, Nicholas Hopper and John Langford, researchers at Carnegie Mellon University. Additional information can be found, e.g., on the webpage \url{http://www.captcha.net}.}) have been widely used by websites to safeguard their services from bot-driven activities. Relatively simple tasks for people, but challenging for computers, are presented in CAPTCHA, such as clicking in a designated area, recognizing letters or numbers that are stretched, selecting objects in an image.
Their simple yet effective mechanisms have contributed to maintain poll integrity, control registration, prevent ticket inflation, and counter false comments, ultimately ensuring a more secure and trustworthy online experience for users.
Image-based CAPTCHAs require both image recognition and semantic classification, making them harder for bots to understand than text-based one. In addition, by collecting data from CAPTCHAs, one could train machine learning models.

It is commonly accepted that human perception and thought are essentially synthetic processes. We immediately form a `Gestalt', or a general idea, of the object. Gestaltic patterns seem to be the primary basis of rational activity as well.  Gestalt-thinking is not well captured by the analytical and compositional framework of classical logical semantics, which essentially deduces the meaning of a composite expression from the meanings of its component parts (`principle of compositionality', see, e.g., \citet{carnap1947}).
For an accurate study of natural languages and creative contexts, where holistic and ambiguous characteristics seem to be significant, classical semantics is no longer very useful. One example in this regard is the last line of Giacomo Leopardi's poem `L'Infinito', ``E 'l naufragar m'è dolce in questo mare'' (And drowning in this sea is sweet to me, i.e. I think it's delightful to dedicate myself to the contemplation and meditation of the infinite). The meanings of the component terms ``naufragar'' (drowning), ``dolce'' (sweet), and ``mare'' (sea) do not correlate to the meanings they have in this context, which appears to be the fundamental cause of the poetic outcome in this instance. By the way, the poem refers to the settlement of Recanati, which is not near a sea. Nonetheless, our words' common meanings continue to be used and are loosely connected to the metaphorical meanings that the poem as a whole evokes, a semantic conundrum that frequently occurs in poetry and musical compositions, where meanings are inherently holistic, contextual and ambiguous. 

Similar issues occur whenever people combine individual concepts to form conceptual combinations or more complex linguistic expressions as sentences and texts. A large empirical literature indeed reveals that the principle of compositionality is systematically violated and, more generally, concepts exhibit aspects of `inherent vagueness', `contextuality' and `emergence' which prevent them from being modelled within classical Boolean logical and classical Kolmogorovian probabilistic structures  (see, e.g., \citet{aerts2009a,aerts2009b,coecke2010,busemeyerbruza2012,aertsbroekaertgaborasozzo2013,aertsgaborasozzo2013,bruzakittorammsitbon2015,dallachiaragiuntininegri2015b}
 and references therein).

Recently, various approaches have been put forward to theoretically cope with this structural inability of classical compositional semantics to handle the dynamics of human concepts. Remarkably, some of these approaches use the logico-mathematical formalism of quantum theory, detached from its physical interpretation, as a modelling tool to represent conceptual meaning. This research fits a growing programme that applies quantum structures in the mathematical modelling of cognitive processes, with relevant extensions to information retrieval processes (see, e.g., \citet{aerts2009a,aerts2009b,coecke2010,busemeyerbruza2012,aertsbroekaertgaborasozzo2013,aertsgaborasozzo2013,bruzakittorammsitbon2015,dallachiaragiuntininegri2015b,pothosbusemeyer2009,frommholzetal2010,khrennikov2010,piwowarskietal2010,bucciomeluccisong2011,ingo2011,havenkhrennikov2013,dallachiaragiuntininegri2015a,kwampleskacbusemeyer2015,melucci2015,aertsarguellesbeltransassolisozzoveloz2017,aertsetal2018,haven2018,pisanosozzo2020,aertssassolisozzoveloz2021} and references therein).

In particular, in the `holistic quantum computational semantics' approach, meanings are typically represented by quantum superpositions of meanings and are treated as fundamentally dynamical objects. Any global meaning determines partial meanings, which are frequently vaguer than the global one. The majority of research on quantum computational logics has focused on sentential logics, whose alphabet is made up of atomic sentences and logical connectives. A first-order epistemic quantum computational logic with a semantic characterization that can express sentences like ``the animal acts,'' ``the animal eats the food'', etc. was presented in \citet{dallachiaraetal2015}. An alternative, but similar, approach considers a concept as an entity whose meaning is incorporated into a given state, whose state can change under the influence of a context, and represents conceptual entities in the formalism of quantum theory in Hilbert space \citep{aertssassolisozzo2016}. Within these two quantum approaches, empirical and theoretical studies have been carried out with the aim of identifying `quantum entanglement' in the combination of natural concepts \citep{bruza2009,bruzakittonelsonmcevoy2009,aertssozzo2011,aertssozzo2014,gronchistrambini2017,arguelles2018,aertsetal2019,beltrangeriente2019,arguellessozzo2020,aertsbeltrangerientesozzo2021,aertsarguellesbeltrangerientesozzo2023b,bertinietal2023}. 

In physics, the notion of entanglement was introduced by Erwin Schr\"odinger in a letter to Albert Einstein, after reading the 1935 article where Einstein, with his two collaborators Boris Podolsky and Nathan Rosen, analysed a specific type of correlations, nowadays known as `EPR correlations', appearing in a composite quantum system, or `entity', made up of two individual entities \citep{epr1935}. In the same year, Schr\"odinger published an article on the matter, where he officially introduced the notion of entanglement as a phenomenon that, while appearing for two quantum entities separated in space, seems to indicate that the entities are `non-separated' \citep{schrodinger1935}. This made Einstein use the expression `spooky action at-a-distance' when talking about the phenomenon. In 1951, David Bohm introduced the archetypical situation of two spin 1/2 quantum entities in the singlet spin state of total spin equal to zero and flying apart \citep{bohm1951}. If one imagines that, (i) if one spin is forced ``up'' by the measurement apparatus applied to it, then, as a consequence, the other spin is `immediately' forced ``down'', even when not measured upon, and (ii) this `mechanism' keeps occurring for two quantum entities independently of their distance, the phenomenon could be called indeed a spooky action at-a-distance. It is John Bell who has the great merit of making the question open to experimental verification by deriving an inequality that should not be violated if certain empirical tests are performed on EPR correlated quantum systems while quantum theory predicts that the inequalities derived by Bell will be violated \citep{bell1964}.
 This has also had a relevance on the so-called `hidden variables programme', initiated after the EPR article, because these results entail that any hidden variable completion of quantum theory has unavoidably to be non-local. 

After the seminal work of Bell, several similar inequalities have been derived, and one typically refers to them as `Bell's inequalities' (see, e.g., \citet{brunner2014}), one of them being the `Clauser--Horne--Shimony--Holt (CHSH) inequality' \citep{chsh1969}. The CHSH inequality is particularly interesting, as it is well suited, not only for empirical tests of non-locality in quantum physics, but also as a test to detect the presence of entanglement, within and beyond physical domains. Indeed, in the absence of entanglement, the CHSH inequality would not be violated.\footnote{We notice that the violation of Bell's inequalities, while being a sufficient condition for the detection of entanglement, is not necessary because, for a given Bell's inequality, entangled states exist which do not violate it.} As a matter of fact, we have performed in the last decade various tests, including text-based cognitive tests on human participants, information retrieval tests on corpora of documents, and image retrieval tests on web search engines, which confirm that the CHSH inequality is systematically violated in cognitive domains, particularly, when two concepts combine \citep{aertssozzo2011,aertssozzo2014,arguelles2018,aertsetal2019,beltrangeriente2019,arguellessozzo2020,aertsbeltrangerientesozzo2021,aertsarguellesbeltrangerientesozzo2023b,bertinietal2023}.

In this article, we deepen and extend the investigation above and present the results of a video-based cognitive test that we have recently performed on the conceptual combination {\it The Animal Acts}, considered as a combination of the concepts {\it Animal} and {\it Acts},\footnote{In this article, we write concepts using capital letters and italics, e.g., {\it Animal}, {\it Fruit}, {\it Vegetables}, etc. This is frequently the way of referring to concepts in cognitive psychology.}  where the term ``acts'' refers to one of the sounds that can be made by an animal. The test used the videos produced by CAPTCHA through artificial intelligence (AI) and showed to the participants, who had to judge among the videos the best examples of {\it The Animal Acts}. As such, the test is more realistic and effective than the previous text-based tests, as it surpasses language barriers and eliminates the need for prior knowledge, thus enabling universal accessibility. We show that the CHSH inequality is significantly violated in the test, which reveals the presence of entanglement between the component concepts. We also work out a quantum theoretical model in Hilbert space for the statistical data (`judgement probabilities'), which reveals that entanglement occurs at both state and measurement levels, hence it is even stronger than the entanglement that is typically detected in quantum physics tests. 

The presence of entanglement can be naturally explained if one observes that both concepts {\it Animal} and {\it Acts} carry meaning, but also the combination {\it The Animal Acts} carries its own meaning, and this meaning is not simply related to the separate meanings of {\it Animal} and {\it Acts} as prescribed by a classical compositional semantics. It also contains emergent meaning almost completely caused by its interaction with the wide overall context. We have called the complex mechanism by which meaning is attributed to the combined concept `contextual updating’, and it occurs at the level of entanglement creation \citep{aertsarguellesbeltrangerientesozzo2023a}.

For the sake of completeness, we summarize the content of this article in the following.

In Section \ref{entanglement}, we illustrate the general setting of a Bell-type test for the empirical detection of entanglement in both physical and cognitive, i.e. conceptual and linguistic, domains. The violation of the CHSH inequality is generally considered as conclusive towards the presence of entanglement in the given situation. We however observe that situations exist in which entanglement cannot only be attributed to the state of a composite entity, but also the measurements need to be considered to be entangled. This typically occurs in cognitive domains.

In Section \ref{marginal}, we explain the general reasons which led us to put forward a theoretical approach that diverges in some aspects from more known approaches.

In Section \ref{test}, we describe the empirical setting of the video-based cognitive test we have performed, stressing the advantages of a cognitive test based on videos with respect to more traditional text-based tests. We also present the empirical results, which violate the CHSH inequality in agreement with those of previous tests, thus indicating that entanglement is a natural candidate to theoretically model the empirical situation. 

In Section \ref{quantum}, we work out a quantum mathematical representation in Hilbert space for the data collected in Section \ref{test} and show that the violation of the CHSH inequality can be attributed to the presence of a strong form of entanglement involving both states and measurements, as anticipated in Section \ref{entanglement}.

In Section \ref{explanation}, we finally provide a theoretical analysis of the results obtained in the test and the ensuing modelling, and establish some deep connections between the non-classical mechanism of meaning attribution and the appearance of entanglement in cognitive domains.

\section{Identifying entanglement in physics and cognition\label{entanglement}}
We present in this section the typical way in which entanglement is theoretically understood and empirically detected in physical and cognitive, i.e. conceptual and linguistic, domains.

In Section 1, we have introduced entanglement as a statistical property of a composite entity, which can be identified by means of the empirical violation of Bell's inequalities \citep{bell1964,brunner2014,chsh1969}. We have also seen that an inequality that is particularly suited for empirical control is the CHSH inequality \citep{chsh1969}. We remind that the violation of Bell's inequalities is typically interpreted by saying that, due to entanglement, micro-physical entities exhibit genuinely non-classical aspects, as non-separability, contextuality and non-locality. We add that the statistical correlations produced in the presence of entanglement cannot be reproduced by any classical model of probability satisfying the axioms of Kolmogorov \citep{accardifedullo1982,pitowsky1989}. Because of its deep physical implications, entanglement has been considered as the distinctive trait of quantum theory since its early days \citep{schrodinger1935}. 

The empirical setting for the identification of entanglement refers to a `Bell-type test' \citep{epr1935,bohm1951,bell1964,brunner2014,chsh1969}, which can be outlined as follows. Let us consider a composite physical entity ${S}_{12}$, prepared in an initial state $p$, and such that the component entities ${S}_1$ and ${S}_2$ can be recognized as component parts of ${S}_{12}$. Next, let us perform the joint measurements $AB$, $AB'$, $A'B$ and $A'B'$ on $S_{12}$, where the joint measurement $XY$ consists in performing the measurement $X$ on $S_1$, with possible outcomes $X_1$ and $X_2$, and the measurement $Y$ on $S_2$, with possible outcomes $Y_1$ and $Y_2$, with $X=A,A'$, $Y=B,B'$. The component entities $S_1$ and $S_2$ have interacted in the past, but are spatially separated when the joint measurements are performed.  If $X_1,X_2,Y_1,Y_2$ can only be $\pm 1$, $X=A,A'$, $Y=B,B'$, the expected values of $AB$, $AB'$, $A'B$ and $A'B'$ are the correlation functions $E(A,B)$, $E(A,B')$, $E(A',B)$ and $E(A',B')$, respectively. One can then prove that, under the reasonable, in classical physics, assumption of `local separability', or `local realism' \citep{epr1935}, the CHSH inequality, namely,
\begin{equation} \label{chsh}
-2 \le E(A',B')+E(A',B)+E(A,B')-E(A,B) \le 2
\end{equation}
should be satisfied \citep{chsh1969}. We call `CHSH factor' the quantity
\begin{equation} \label{chshfactor}
\Delta_{CHSH}=E(A',B')+E(A',B)+E(A,B')-E(A,B)
\end{equation}
and observe that $\Delta_{CHSH}$ is mathematically bound by $-4$ and $+4$.

According to modern manuals of quantum theory, the component entities $S_1$ and $S_2$ are associated with the complex Hilbert spaces ${\mathscr H}_1$ and ${\mathscr H}_2$, respectively. In this case, both ${\mathscr H}_1$ and ${\mathscr H}_2$ are isomorphic to the complex Hilbert space $\mathbb{C}^2$ of all ordered pairs of complex numbers. The composite entity $S_{12}$ is instead associated with the tensor product Hilbert space ${\mathscr H_1}\otimes {\mathscr H}_2$. In this case, ${\mathscr H_1}\otimes {\mathscr H}_2$ is isomorphic to the complex Hilbert space $\mathbb{C}^2 \otimes \mathbb{C}^2$. The possible (pure) states of $S_1$ and $S_2$ are represented by unit vectors of ${\mathscr H}_1$ and ${\mathscr H}_2$, respectively, and the measurements that can be performed on $S_1$ and $S_2$ are represented by self-adjoint operators on ${\mathscr H_1}$ and ${\mathscr H}_2$, respectively. However, ${\mathscr H_1}\otimes {\mathscr H}_2$ also contains vectors that cannot be written as the tensor product of a unit vector of ${\mathscr H_1}$ and a unit vector of ${\mathscr H_2}$. These non-product vectors of ${\mathscr H_1}\otimes {\mathscr H}_2$ are said to represent `non-product', or `entangled', states of $S_{12}$. Analogously, the self-adjoint operators of ${\mathscr H_1}\otimes {\mathscr H}_2$ are not limited to operators that are the tensor product of a self-adjoint operator of ${\mathscr H}_1$ and a self-adjoint operator of ${\mathscr H}_2$. In these cases, at least one eigenvector of these non-product self-adjoint operators represents an entangled state. These non-product self-adjoint operators are said to represent `non-product', or `entangled', measurements of $S_{12}$ \citep{brunner2014}. 

The CHSH inequality in Equation (\ref{chsh}) is manifestly violated in quantum theory and, when a violation occurs, it is due to the presence of entanglement between the component entities $S_1$ and $S_2$ that can be recognized as component parts of $S_{12}$. Equivalently, in the absence of entanglement, the inequality in Equation (\ref{chsh}) would not be violated. As we have anticipated in Section \ref{intro}, the typical situation in which the inequality in Equation ({\ref{chsh}) is violated in quantum theory consists in the state $p$ of the composite entity $S_{12}$ being the singlet spin state (an example of a maximally entangled state \citep{brunner2014}) and the joint measurements $XY$ being product measurements, $X=A,A'$, $B'B'$. This situation entails a CHSH factor equal to $\Delta_{QMC}=2\sqrt{2} \approx 2.83$, known as `Cirel'son's bound'  \citep{cirelson1980,cirelson1993}. A situation involving product measurements automatically satisfies the `marginal law of Kolmogorovian probability', i.e. the conditions that, for every $i,j=1,2$, 
\begin{eqnarray}
\sum_{j=1,2} \mu(X_iY_j)&=&\sum_{j=1,2}\mu (X_iY'_{j}) \label{marginal1} \\
\sum_{i=1,2} \mu (X_iY_j)&=&\sum_{i=1,2}\mu (X'_iY_{j}) \label{marginal2}
\end{eqnarray}
In Equation (\ref{marginal1}), $\mu(X_iY_j)$ ($\mu(X_iY'_j)$)  is the joint probability of obtaining the outcomes $X_i$ in a measurement of $X$ on $S_1$ and $Y_j$ ($Y'_j$) in a measurement of $Y$ ($Y'$) on $S_2$, $X=A,A'$, $Y,Y'=B,B'$, $Y' \ne Y$. In Equation (\ref{marginal2}), $\mu(X_iY_j)$ ($\mu(X'_iY_j)$)  is the joint probability of obtaining the outcomes $X_i$ ($X'_i$) in a measurement of $X$ ($X'$) on $S_1$ and $Y_j$ in a measurement of $Y$ ($Y'$) on $S_2$, $X,X'=A,A'$ and $Y=B,B'$, $X' \ne X$. The conditions in Equations (\ref{marginal1}) and (\ref{marginal2}) are also known as `no-signaling', or `marginal selectivity', conditions, and have been widely studied in both physical (see, e.g., \citet{brunner2014,peresterno2004,horodecki2009}) and cognitive literature (see, e.g., \citet{barros2015,dzhafarov2016}). A violation of these conditions has been traditionally considered as problematical by physicists, as it would open up the possibility of a faster-than-light communication. We will come back to this point in Section \ref{marginal}.

The needs of quantum computation and quantum information have intensified the theoretical research on Bell's inequalities (see, e.g., \citet{brunner2014,genovese2005}) as well as the research on the empirical consequences of entanglement (see, e.g., \citet{genovese2005,vienna2013,urbana2013}), confirming the predictions of quantum theory.

In the meanwhile, it has become evident that the genuinely quantum aspects of contextuality, entanglement, indistinguishability, interference, and superposition, are not peculiar of micro-physical entities, as they also occur in the quantum mathematical modelling of complex cognitive processes, e.g., the processes involving categorization, judgement, decision, perception, and language. In particular, both experimental and theoretical research has been dedicated to the identification of entanglement in the combination of natural concepts, and our research team has provided substantial contributions along both lines of this investigation, as follows.

At an experimental level, we performed cognitive tests on human participants \citep{aertssozzo2011,aertssozzo2014,aertsarguellesbeltrangerientesozzo2023b,bertinietal2023}, document retrieval tests on structured corpuses of documents \citep{beltrangeriente2019,aertsbeltrangerientesozzo2021} and image retrieval tests on web search engines \citep{arguelles2018,arguellessozzo2020,bertinietal2023} using various combinations of two concepts. We mainly investigated the conceptual combination {\it The Animal Acts}, which we considered as a composite conceptual entity made up of the individual conceptual entities {\it Animal} and {\it Acts}. The tests had the form of the Bell-type test above, and we found (i) a systematic violation of the CHSH inequality in Equation (\ref{chsh}). However, we also found (ii) a systematic violation of the marginal law conditions in Equations (\ref{marginal1}) and (\ref{marginal2}) and, in some cases, (iii) the CHSH factor in Equation (\ref{chshfactor}) exceeded Cirel'son's bound.

While empirical finding (i) substantially agreed with the predictions of quantum theory and indicated the presence of `conceptual entanglement', empirical findings (ii) and (iii) were unexpected, as they are not believed to occur in quantum physics. This led us to initiate a theoretical investigation on a problem that is usually overlooked in physics, the `identification  problem', that is, the problem of recognising individual entities of a composite entity by performing on the latter the typical joint measurements of Bell-type tests \citep{aertssozzo2014}. We thus elaborated a general theoretical framework to model any Bell-type situation, independently of the nature, physical or cognitive, of the entities involved, within the mathematical formalism of quantum theory in Hilbert space \citep{aertssozzo2014,aertsetal2019}. In this theoretical framework, one applies the standard prescription of quantum theory according to which the composite entity $S_{12}$ needs to be associated with a complex Hilbert space whose dimension is determined by the number of distinct outcomes of the performed measurements. Since in principle, each of the joint measurements 
$AB$, $AB'$, $A'B$ and $A'B'$ have four distinct outcomes, $S_{12}$ should be represented in the Hilbert space $\mathbb{C}^{4}$ of all ordered 4-tuples of complex numbers. Only in the attempt of `recognizing' individual entities $S_1$ and $S_2$ within the composite entity, one considers possible isomorphisms with the tensor product Hilbert space $\mathbb{C}^{2} \otimes \mathbb{C}^{2}$, where each copy of $\mathbb{C}^{2}$ takes into account the fact that measurements with two distinct outcomes can be performed on $S_1$ and $S_2$ in a Bell-type setting. And it is only at this stage, i.e. when individual entities are recognized from measurements performed on the composite entity, that entanglement can be identified. We proved in \citet{aertssozzo2014}, that no unique isomorphism exists between $\mathbb{C}^{4}$ and $\mathbb{C}^{2} \otimes \mathbb{C}^{2}$, and this is the reason why, from a mathematical point of view, different ways exist to account for entanglement being present within the composite entity $S_{12}$ with respect to the individual entities $S_1$ and $S_2$ that are recognized as parts of $S_{12}$. 

Essentially, entanglement manifests itself when the probabilities of a joint measurement on $S_{12}$ cannot be written as products of probabilities of measurements on the component entities $S_1$ and $S_2$. Hence, entanglement is a property of the relation between the joint measurements and measurements on the component entities. Only when the additional symmetry mentioned above, i.e. the marginal law conditions being satisfied, is present in all joint measurements, the entanglement of these different joint measurements can be captured in a state of the composite entity. If this is true for all joint measurements, one can prove that there is only one isomorphism connecting $\mathbb{C}^{4}$ with $\mathbb{C}^{2} \otimes \mathbb{C}^{2}$, and $\mathbb{C}^{4}$ can be directly replaced by $\mathbb{C}^{2} \otimes \mathbb{C}^{2}$ in that case. This means that the situation usually reported in modern manuals of quantum theory is exceptional and not the general one \citep{aertssozzo2014}. In this general situation, where the marginal law conditions are empirically violated, no unique isomorphism exists between $\mathbb{C}^{4}$ and $\mathbb{C}^{2} \otimes \mathbb{C}^{2}$, hence entanglement cannot by captured only by the state. As a matter of fact, empirical violations of Equations (\ref{marginal1}) and (\ref{marginal2}) have been identified in cognitive domains, as anticipated above, but also in physical domains \citep{adenierkhrennikov2007,adenierkhrennikov2016,bednorz2017,deraedt2012,kupczynski2017}. So far, however, relatively little attention in physics has been paid to the violation of the marginal law conditions in physical tests on entangled quantum quantities, and usually the explanation is sought in artifacts of the measurement process \citep{adenierkhrennikov2016}. Because our research group has been one of the first to encounter empirical violations of the CHSH inequality in cognition and, at the same time, a violation of the marginal law conditions \citep{aertssozzo2011}, our attention has been focused on the presence of both violations even in these early years, which includes situations where such violation appeared in tests on entangled physical entities \citep{adenierkhrennikov2007}.

Thus, we devote the next section to outline the rationale underlying our approach to entanglement and its novelties with respect to more known approaches.

\section{The novelties of our approach to entanglement\label{marginal}}
Let us outline how our research on the presence of entanglement in human cognition has influenced our general way of considering entanglement as a phenomenon, including how we consider the phenomenon in its physical presence (for a more detailed account of this, we refer to our various published articles on this topic, e.g., \citet{aertssozzo2014,aertsetal2019}). Since the experimental situations we considered concern bipartite entities consisting of concepts that are combined and not physical entities that can be localized in space, as is the case for the study of entanglement in physics, we found it desirable to switch to a deeper characterization of the phenomenon concerning its non-classicality. More precisely, it should be a characterization that does not depend on the presence in space of the entities under consideration. This means that the notion usually associated with entanglement when the phenomenon occurs in physics, namely, `non-locality', is not the notion desirably to use to characterize entanglement in the first place. More, even also the connection of the marginal law conditions with no-signaling should best be handled with caution in that respect, because there too the emphasis is on two entities that exist separately and, in some unspecified way, are each of them localized in space, often personificated by Alice and Bob, who can exchange signals with each other. The unconsciously powerful image thereby projected onto a situation of entanglement should not be underestimated in the already too narrow limitation it imposes on what entanglement really can be. Therefore, we prefer not to simply connect the situation in which the marginal law of Kolmogorovian probability is violated in an indiscriminate way with the physical imagined situation of no-signaling between Alice and Bob. We indeed have now become quite convinced, mainly through our research in quantum cognition, that entanglement confronts us with something `holistically deeper', both when it occurs in human cognition and when it occurs in nature. Fortunately, there are plenty of notions that are indeed present in a well-defined way, and can be introduced mathematically rigorously in both physical and cognitive situations, that make it possible to characterize entanglement without taking recourse to an image which would stand in the way for a view that is closer to the nature of entanglement. Let us first bring out concretely the most important of these elements before returning to specifying how, in our opinion, we can also bring the image of entanglement closer to its deep nature.  

In both physical and cognitive realms, the entities studied consist of two component entities. Although we have to immediately nuance this aspect, namely, that we do not know exactly in which way the two component entities are connected to each other and to the entity which is composed of them, finding out that `is', by the way, the principle research aim on the phenomenon of entanglement, as originally pointed out already by Schr\"odinger. We have called such entities `bipartite entities'. 

Probabilities are rigorously defined as limits of relative frequencies in both physical and cognitive realms, and this is where the first aspect of the characterization of entanglement comes in. There is entanglement if the probabilities obtained from the joint measurements on the bipartite entity are not a factorization of the probabilities obtained from the separate measurements that make up the joint measurement, the bipartite entity being prepared in a well-defined state. Thus, this means that we consider the notions of `joint measurement' and `bipartite entity prepared in a well-defined state' to be elements needed to characterize entanglement, and indeed, for the cognitive situations we studied, we have always taken care of these elements to be well-defined. 

What becomes the `non-classicality' of this general characterization of entanglement? Well, if classical physical entities are separated from each other in space, then the aforementioned probabilities factorize. And this archetypical situation that we know well, even from our everyday experience with classical entities, is why notions of non-locality and no-signaling made their appearance in the study of entanglement. Also, when bipartite quantum entities described by the tensor product procedure are in a product state, and the joint measurements are represented by product operators, then the said probabilities factorize. That means that in a situation modelled with the quantum formalism, if these probabilities do not factorize, at least one of these aspect fails. The most studied failing aspect is when the state of the bipartite entity is not a product state, but a superposition of product states, which is customarily now called an entangled state. Little attention was paid to the situation where the joint measurements are not represented by product operators, which also leads to a situation where the probabilities do not factorize. In what follows we will analyze why this is the case and why our approach therein differs from the most common one.

What is the status of the violation of the CHSH inequality within the more general framework in which we want to study both physical and cognitive situations with entanglement? Here we use Pitowsky's study of correlation polytopes and their connection with classical Kolmogorovian probabilities \citep{pitowsky1989}. This theory proves that satisfying the CHSH inequality in all its variants is equivalent to the existence of a classical Kolmogorovian probability model for the corresponding probabilities. Pytowsky's theory thus shows that the violation of one of the CHSH inequalities is sufficient for the presence of a non-classical probability structure. This does not mean that it must necessarily be a quantum probability structure, but it does justify the motivation to try quantum probability as a modelling tool, which is what we did in our research.

Little attention has been paid to the situation where entanglement has its origin in violating both the CHSH inequality and the marginal law conditions. The reason is that violating the marginal law conditions which is equivalent to violating the no-signaling conditions is by many interpreted as indicating the presence of signaling, and therefore these situations are misidentified by many physicists as situations where signaling is present, and thus trivially uninteresting in terms of the phenomenon of entanglement. What is the error in the reasoning made here? The no-signaling conditions only constitute a `sufficient' condition for no-signaling to be present, but it is not a `necessary' condition. From a logical perspective, this is a rather subtle situation where typically logical reasoning errors are made, and we believe this is frequently the case here. Thus, we stress that, even if the no-signaling conditions are not satisfied, it is perfectly possible that no signals (can) be sent between Alice and Bob.

Let us now bring out the elements that motivate us to interpret the violation of the marginal law conditions as being due to a lack of presence of some form of symmetry in the considered situation rather than as a consequence of the existence of signaling.

Already in the early days of studying entanglement, around the same period of Aspect's experiments, one of the authors of this article proposed an explicit mechanism that is abundantly present in nature (and later identified in human cognition) in a subtle but obvious way, and that violates Bell's inequalities \citep{aerts1982,aerts1983}. The mechanism starts from the mere potential presence of correlations in the state of the bipartite entity -- in the example worked out in \citet{aerts1982,aerts1983}, this potential presence is the state of water contained in two vessels connected by a tube, and that potential is actualized by and during the joint measurements themselves -- in the example in \citet{aerts1983}, the water volumes extracted from each vessel by siphons will end up  correlated when collected in outcome reference vessels due to the connecting tube and the classical principle of communicating vessels (see Figure \ref{vessels}). The group of scientists studying the phenomenon of entanglement in those early days was tiny and without the authority to deal with anything important. As a consequence, all sorts of hypotheses were considered. Even the hypothesis that quantum theory may have wrongly predicted a violation of Bell's inequalities was among the opinions of quite a few members of that little group, including Bell himself, but also Clauser's motivation was to find where quantum theory was wrong. Things turned out differently and entanglement showed finally to be a phenomenon real in all its aspects predicted by quantum theory. The common way of looking at entanglement in those days, expressed by Einstein as `spooky action at-a-distance', seemed to carry an irreconcilable conflict with relativity theory. Meanwhile, however, the no-signaling conditions and their structural equivalence with the marginal law conditions also became known, hence entanglement could not be used to send signals between Alice and Bob.
\begin{figure}
	\begin{center}
		\includegraphics[scale=0.18]{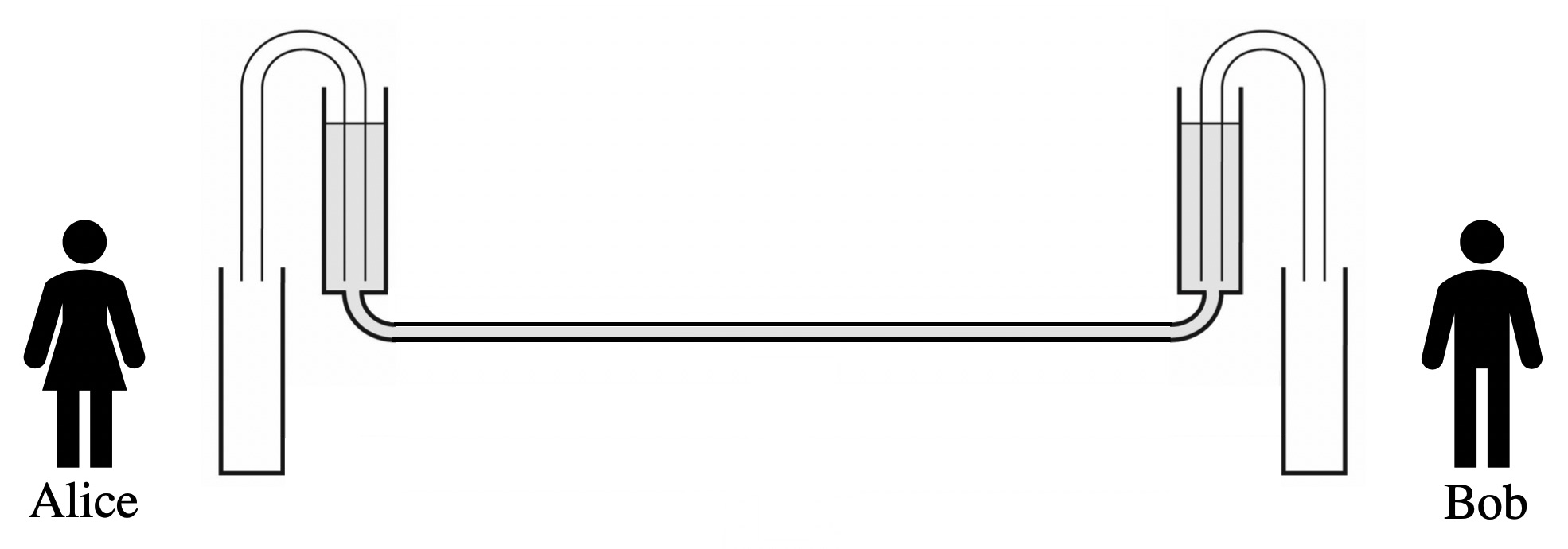}
	\end{center}
	\caption{A schematic representation of the vessels of water example worked out in \citet{aerts1982,aerts1983} (see also the discussion in \citet{aertssassolidb2021}).
	\label{vessels}}
\end{figure}
The example of a violation of Bell's inequalities from the presence of potential correlations only actualizing pduring the joint  measurements proposed in \citet{aerts1982,aerts1983} also violates the no-signaling conditions. Therefore, later on, when in the 1990s a much larger group of physicists and computer scientists started analyzing quantum structures, mainly out of interest in quantum computation, and somewhat later also quantum information, it was classified as a mechanism that is not really interesting, since identified as violating Bell's inequalities because of containing signaling.  However, However, later, variants of the same mechanism were created where the no-signaling conditions are not violated \citep{aerts1991,aerts2005} and, inspired by this, an example was also worked out in human cognition, with a combination of concepts, where the same mechanism causes a violation of Bell's inequalities while satisfying the no-signaling conditions \citep{aerts2014}. If these examples are compared, considering the proposed mechanism in detail, it is clear that whether or not the no-signaling conditions are satisfied is mainly a matter of presence of enough symmetry in the mechanism, but that the mechanism, even in the cases where due to absence of the necessary symmetry the no-signaling conditions are violated, contains no-signaling nevertheless. We hereby recall again that the no-signaling conditions only constitute a sufficient condition in connection with no-signaling but not a necessary condition, as it is often considered erroneously to be.

A second element related to our interpretation of the violation of the marginal law conditions in most experiments in human cognition that violate Bell's inequalities, as being due to a `lack of symmetry' and not to the presence of signaling, consists in a reappraisal of the mechanism itself. More than before, we have become convinced that the mechanism proposed in the various examples \citet{aerts1982,aerts1983,aerts1991,aerts2005,aerts2014}, is also the mechanism that violates Bell's inequalities in physics and let us make the analysis for the case of the singlet spin state that supports us in this conviction. The singlet spin state is a superposition of two product states, associated with the pairs (up, down) and (down, up) of the respective spins of the two component entities of the composite bipartite entity, where the (up, down) and (down, up) can be in any possible space direction and its opposite. That means it is a state that only potentially holds correlations, where these correlations are actualized during the spin measurements. This is exactly the mechanism by which the examples above were constructed, and the macroscopic model in \citet{aerts1991} realizes the exact probabilities and states within two coupled Bloch spheres. By the way, we can also understand after this analysis why no spooky action at-a-distance takes place. The collapse as a result of the measurement happens indeed from the singlet spin state as superposition to one of those product states, (up, down) or (down, up) in some space direction, even if meanwhile the entities carrying the spin are far away from each other in space. The collapse is in that sense a change of state which, however, is not at all connected with whether or not the entities are far apart in space, and it gives rise to outcome events that are space-like separated. Also for the examples elaborated in \citet{aerts1982,aerts1983,aerts2005,aerts2014}, the distance between the outcome regions does not play any role in the dynamics that plays out for the outcome to occur and the outcomes are also events that are space-like separated. The examples in \citet{aerts1982,aerts1983,aerts1991,aerts2005,aerts2014} are macroscopic and do not contain microscopic quantum elements, does this mean that we believe entanglement also admits a macroscopic variant? Our answer to this question is positive with the nuance that, as is the case for microscopic realizations of entanglement, a correlation coherence region comes into play, where it is the quality of the experimental setup that should ensure that this correlation coherence region is preserved during the experimental test and is large enough to contain the outcome regions. It should not be surprising that in the macroscopic realm with rather classical entities, it is possible to realize a `potentiality for correlations', think of how potential energy in the same classical environment realizes a `potentiality for energy'. But also note that for the macroscopic physical examples it is necessary that there remains a connection in space between the two part entities which is not necessary for the quantum microscopic examples, and one could say that this is the way that non-locality manifests itself in our approach. We can see for this aspect the connection with a recent perspective on entanglement that we developed, where it may have its origin in human cooperation to combat uncertainty, not occurring spontaneously in that sense in the macroscopic world, which may be a difference from the more spontaneous way it occurs in the microscopic quantum world \citep{aertsarguellesbeltrangerientesozzo2023a}. 

There remains one more element that needs explanation, namely, how it is that the size of the CHSH inequality expression for cognitive tests, and likewise for the macroscopic examples, can exceed Cirel'son's bound, while theoretically this is not the case for microscopic quantum entities that are entangled. We can understand the condition for the existence of this bound when we consider Cirel'son's proof of it in detail. We can then note that it is necessary to be able to represent the four considered joint measurements present in the CHSH inequality expression by one self-adjoint operator in the considered Hilbert space. This indicates the supposed presence of a very large and rigorous quantum coherence that incorporates the four joint measurements simultaneously and independently of how and when they may be performed separately. This very large internal coherence is not easy to bring present in the cognitive tests, nor in the macroscopic tests, although it was realized in \citet{aerts1991}, which gives rise to identical probabilities as the spin of a spin 1/2 quantum entity. The other examples violate Cirel'son's bound, and this is probably linked to the same lack of symmetry present, which in most cognitive tests, and in the most obvious macroscopic examples, also violates the marginal law conditions.

It is already mentioned that our investigation of the phenomenon of entanglement in quantum cognition made us realize that something of a deeper holistic nature is going on when two entities enter into a situation of entanglement with each other. We have now brought out all the elements to make this also explicit. The entangled entity is more than usually supposed to be a new entity in itself only still connected to the original entities from which it was formed in a very specific way, and this has become even more clear to us by analyzing and understanding more and more deeply how concepts are entangled in human language. We recently introduced a new expression `contextual updating' taking place in human language relative to the global meaning carried by the whole context, which contains an important part of our new understanding. That also the marginal law conditions are violated as a consequence of the entangled entity behaving fully as a new entity relative to the global meaning context is clearly identified in language, and it is also clearly seen that this has nothing to do with the presence of signals. We believe that a similar process takes place when two physical entities transform into an entangled entity, namely, a new entity is created that forms itself not primarily with respect to the two component entities, but contextually with respect to the global quantum coherence present, in which the two component entities will generally play an important role, but in principle not only they \citep{aertsarguellesbeltrangerientesozzo2023a} (see Section \ref{explanation}). In this sense, we think that the situation where both the CHSH inequality and the marginal law conditions, are violated is the default situation in terms of entanglement, namely, the situation where its intrinsic reality is best expressed. Admittedly, it remains important to verify that there is no-signalling present that causes the violations, if one wants to distinguish the cases of non-signaling entanglement from the cases where the entanglement is provoked by signaling. We remind however that, even when signaling is involved, a non-classical probability is generated by it, and the phenomenon remains non-classical.

The considerations above led us to investigate in detail the above mentioned identification problem and to elaborate a quantum mathematical representation of the Bell-type tests that violate both the marginal law conditions, the CHSH inequality and Cirel'son's bound. In this quantum theoretical framework, we explicitly introduce entangled it measurements \citep{aertssozzo2014,arguellessozzo2020,aertsbeltrangerientesozzo2021,aertsarguellesbeltrangerientesozzo2023b}. Specifically, we have proved that, whenever the concepts {\it Animal} and {\it Acts} combine to form the combination {\it The Animal Acts}, a strong form of entanglement is created between {\it Animal} and {\it Acts}, which is such that, not only the state of the composite entity {\it The Animal Acts} is entangled, but also the joint measurements are entangled.

We will see in Sections \ref{quantum} and \ref{explanation} that the entanglement above is due to the peculiar way in which the meaning of {\it The Animal Acts} relates to the meanings of {\it Animal} and {\it Acts}, which violates the classical semantic rules of composition. Before doing this, however, we need to present the details of the novel cognitive test based on videos which we have recently performed on this conceptual combination. This will be the aim of Section \ref{test}.

\section{A novel video-based cognitive test\label{test}}

We present in this section the details of the video-based cognitive test on human participants for the detection of entanglement in the conceptual combination {\it The Animal Acts}.  

As anticipated in Sections \ref{intro} and \ref{entanglement}, we consider the concept {\it The Animal Acts} as a combination of the individual concepts {\it Animal} and {\it Acts}, where by ``acts'' we mean the action of producing a recognizable sound by the animal. Next, we consider two pairs of items of {\it Animal}, namely, ({\it Horse}, {\it Bear}) and ({\it Tiger}, {\it Cat}), and two pairs of items of {\it Acts}, namely, ({\it Growls}, {\it Whinnies}) and ({\it Snorts}, {\it Meows}). We are now ready to illustrate the test.

A sample of 221 individuals were presented in a `within subjects design' a HTML5 questionnaire which contained four joint measurements $AB$, $AB'$, $A'B$ and $A'B'$ whose setting was similar to the typical setting of a Bell-type test presented in Section \ref{entanglement}. More specifically, participants were preliminarily asked to read an `introductory text' where an explanation of the type of judgement test they had to complete and a description of the tasks involved in the judgement test were provided. More specifically, the participants had to preliminary complete a simple introductory test on the concept {\it Fruit}, judging the item that they considered as a good example of the concept {\it Fruit}. The items of {\it Fruit}, indicated in Figure \ref{fig_fruit}, were generated by image-based generative intelligence, and we proposed the item {\it Strawberry}, which is an enlarged portion of an inflorescence, and  {\it Tomato}, which is a fruit, to draw maximum attention to the choice. {\it Cherry} and {\it Pomegranate} were the remaining items of {\it Fruit} to judge upon.
\begin{figure}
	\begin{center}
		\includegraphics[height=3 cm]{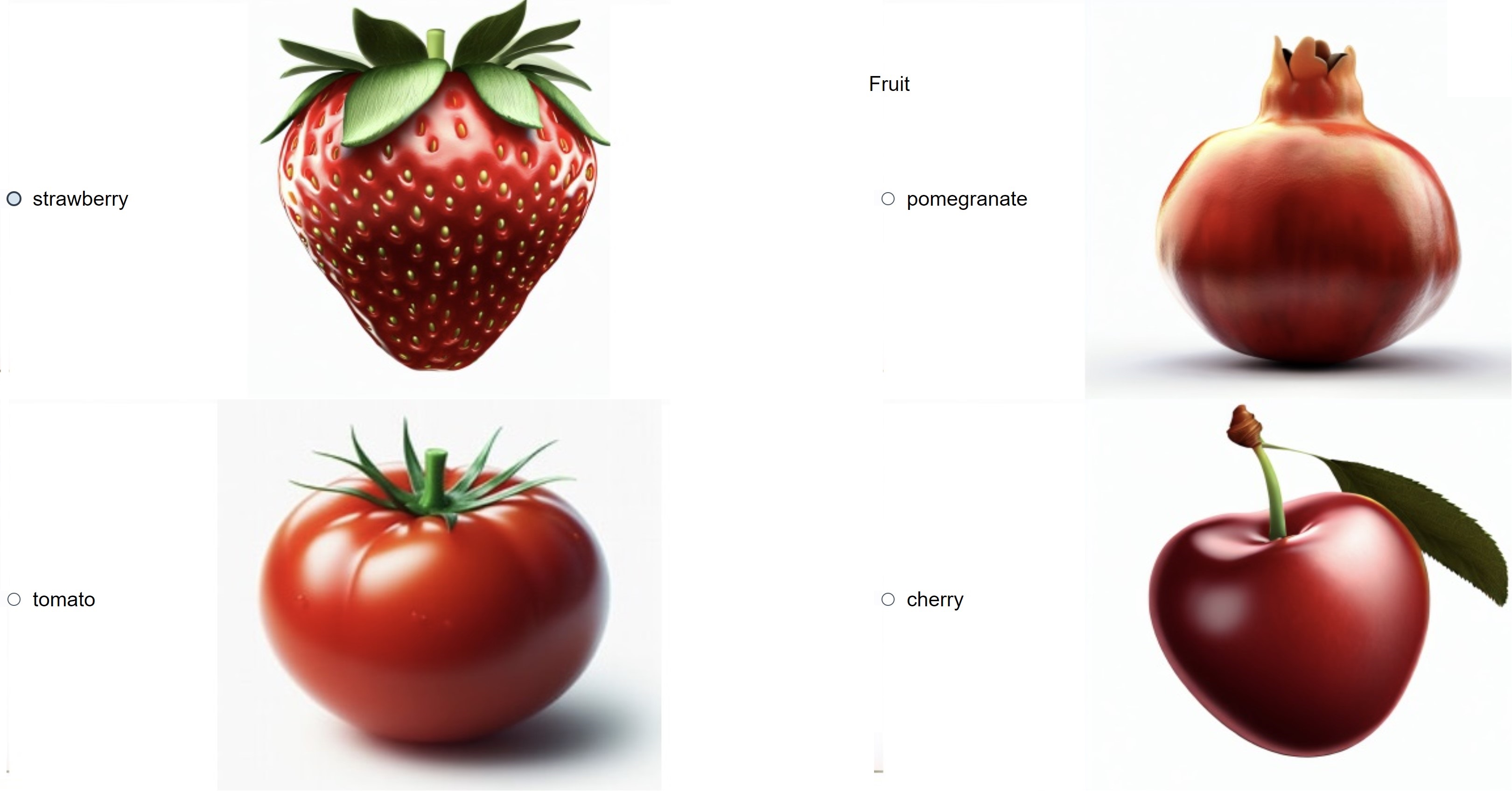}
	\end{center}
	\caption{Introductory test based on different items of the concept {\it Fruit}.
	\label{fig_fruit}}
\end{figure}
Then, in each joint measurement, participants were asked to choose which item in a list of four items they judged as a good example of the conceptual combination {\it The Animal Acts}. In each joint measurement, each of the four possible items of {\it The Animal Act} was a video showing an animal and the corresponding action. More specifically, the videos, indicated in Figure \ref{fig_animal_acts}, were created as follows:

(i) using a video search engine to find a video of the animal producing the sound;

(ii) extracting a cropped portion of the video where the animal produces the sound;

(iii) in cases where the animal and sound combination is unlikely, e.g., a cat that whinnies, using reverse image search to find a similar video of the animal instead of using AI services, as Sora AI, or morphing techniques to create a video of the animal producing the sound.

All of the operations above can be automated by an AI service. Future AI advancements will allow for the creation of 3D animated models that are virtually indistinguishable from real 3D video which can be activated by gaze and selected with pinch gestures, as in Vision Pro and Quest Pro, to create more engaging CAPTCHAs.
\begin{figure}
	\addtocounter{figure}{-4}
	\begin{center}
		\begin{minipage}{0.25\linewidth}
			\includegraphics[height=2 cm]{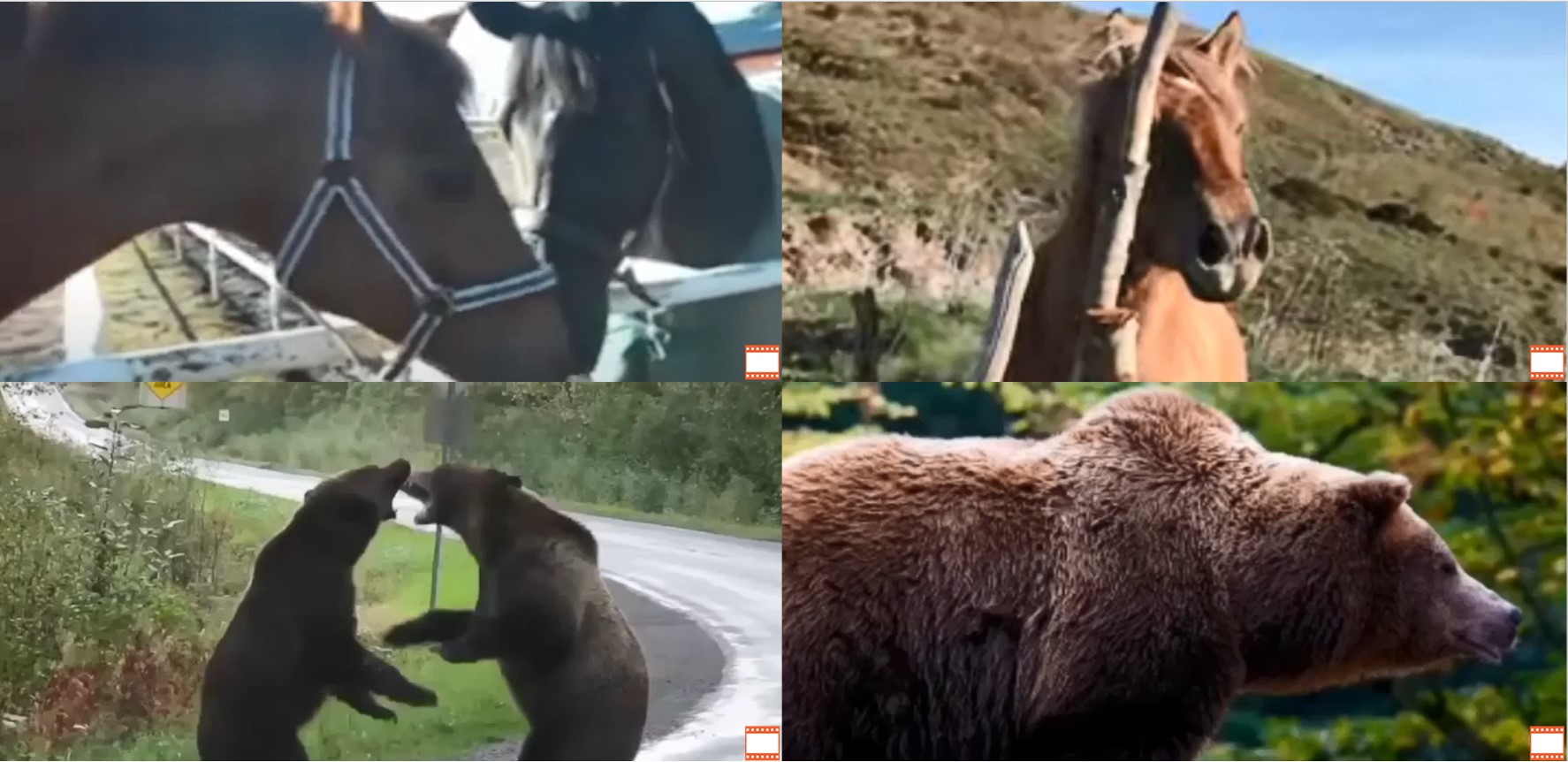}
			\captionsetup{labelformat=empty}
			\captionof{figure}{sub-test 1}
		\end{minipage}%
		\begin{minipage}{0.25\linewidth}
			\includegraphics[height=2 cm]{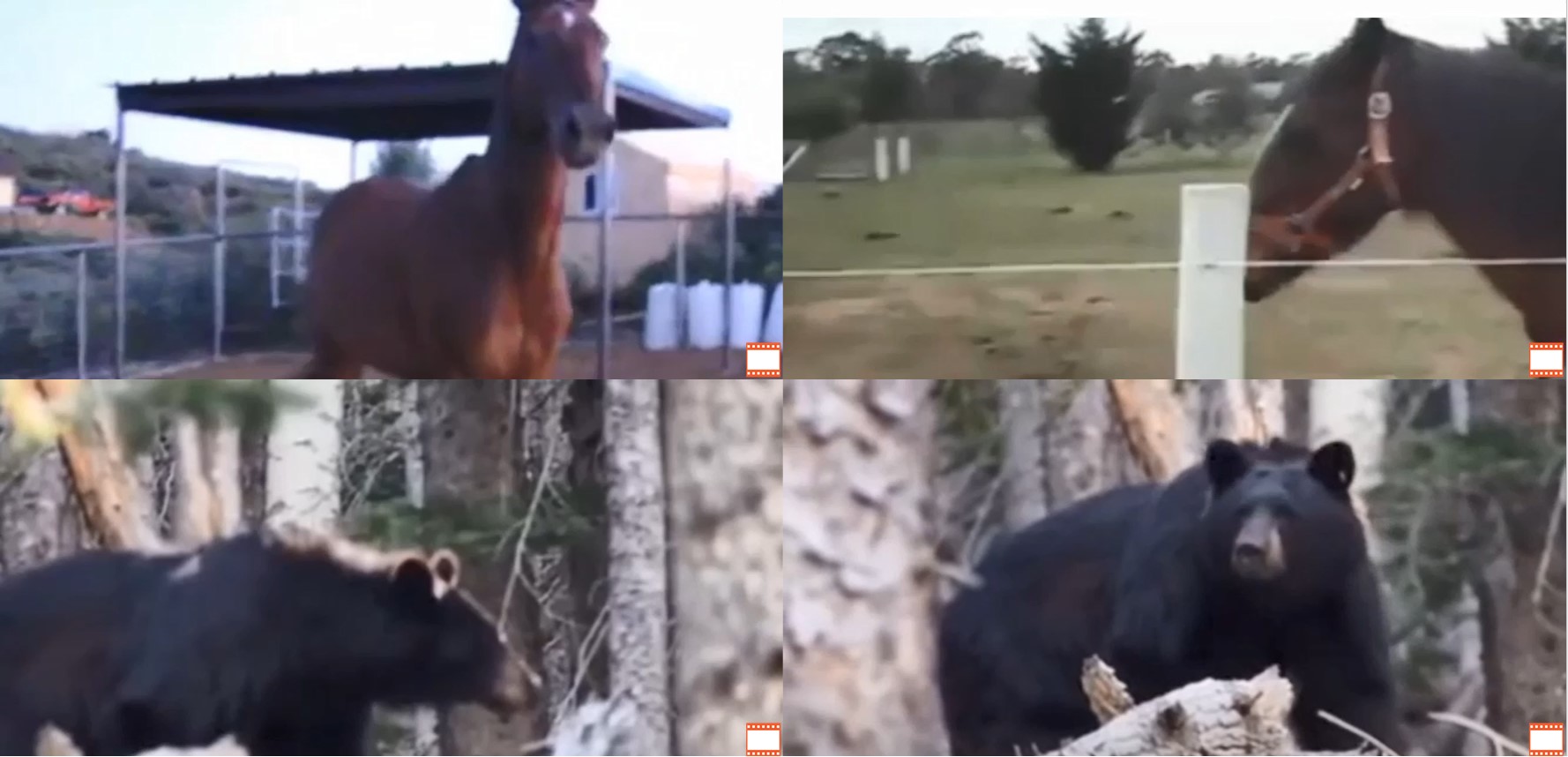}
			\captionsetup{labelformat=empty}
			\captionof{figure}{sub-test 2}
		\end{minipage}%
		\begin{minipage}{0.25\linewidth}
			\includegraphics[height=2 cm]{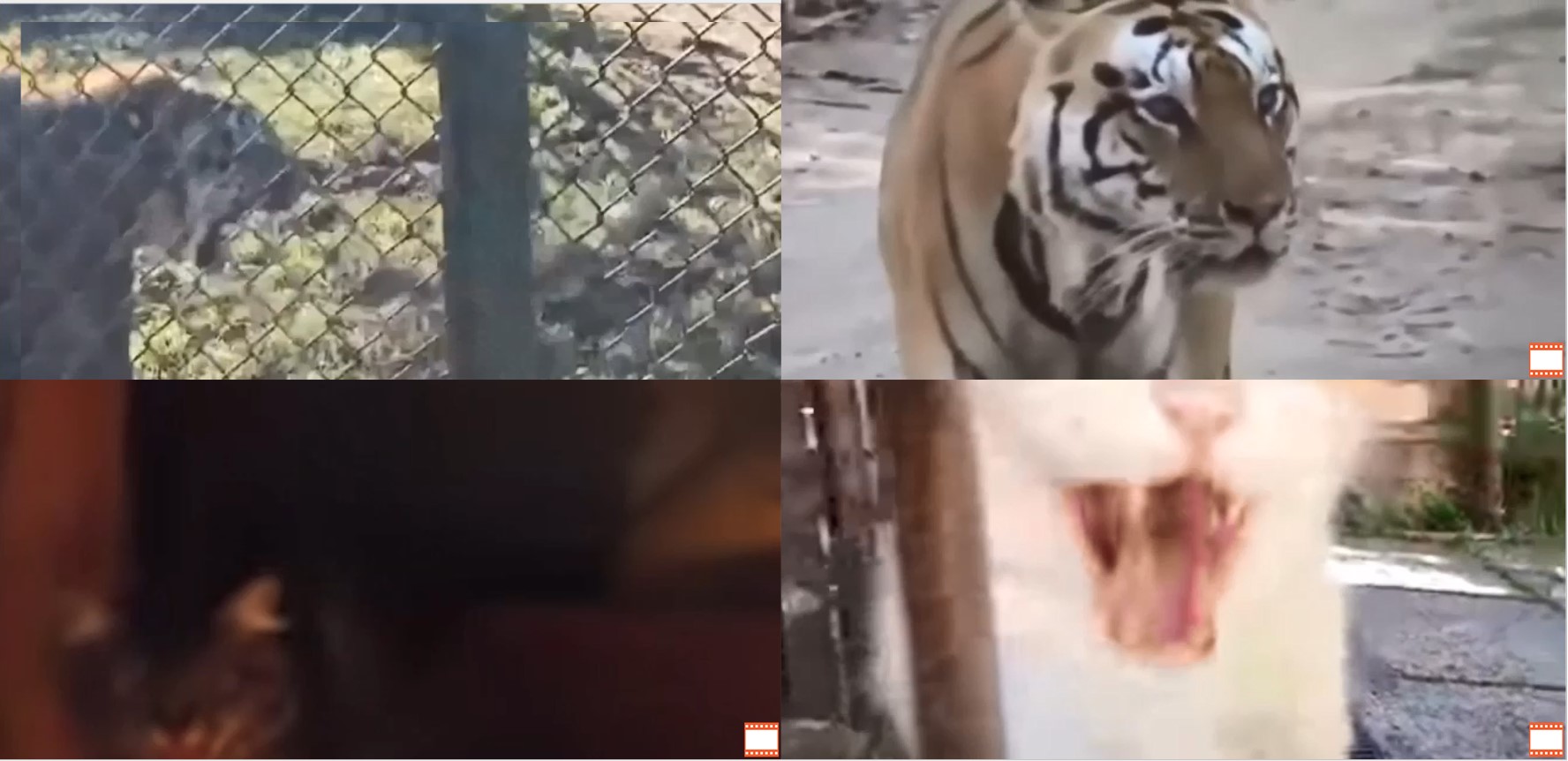}
			\captionsetup{labelformat=empty}
			\captionof{figure}{sub-test 3}
		\end{minipage}%
		\begin{minipage}{0.25\linewidth}
			\includegraphics[height=2 cm]{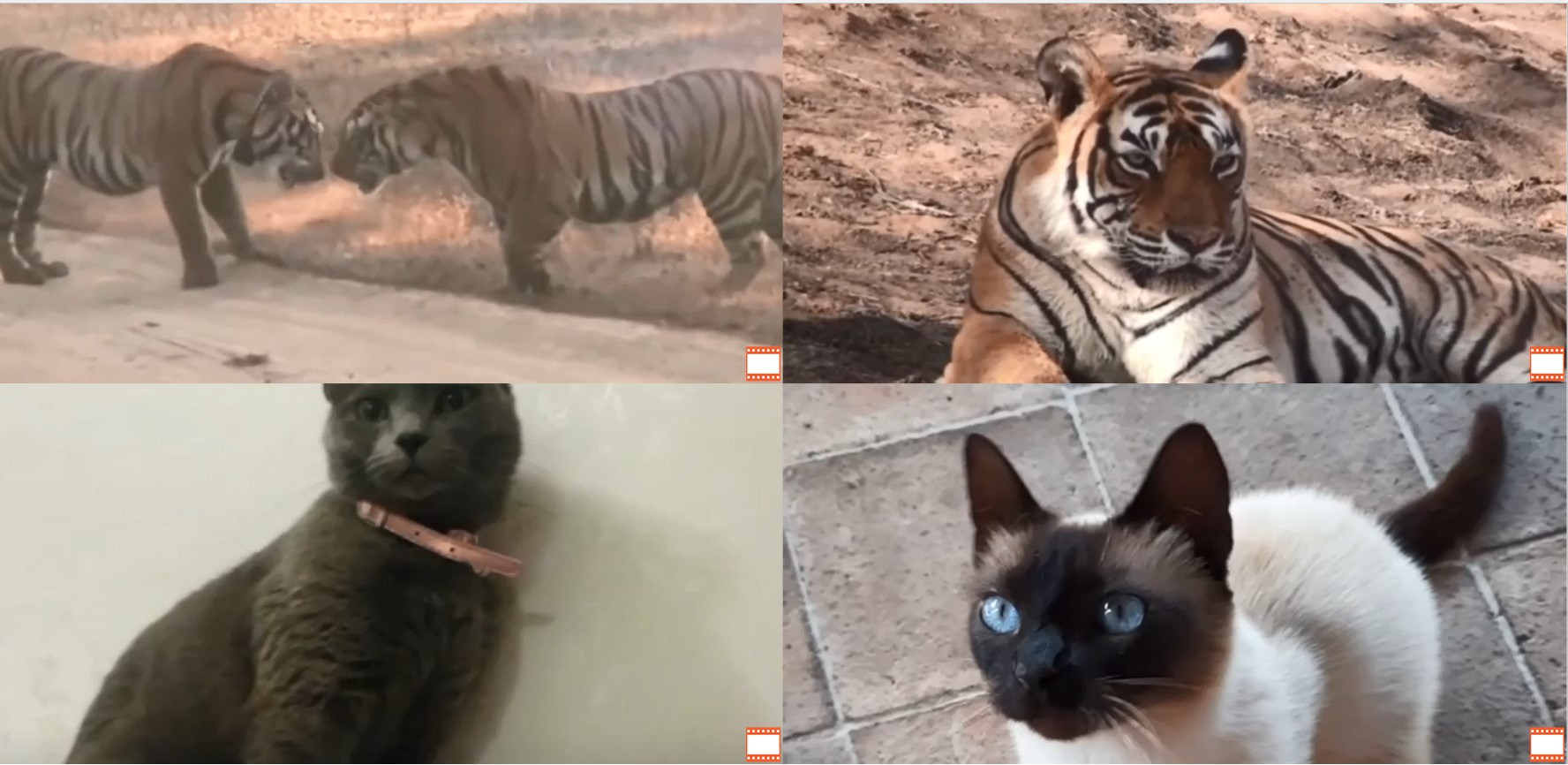}
			\captionsetup{labelformat=empty}
			\captionof{figure}{sub-test 4}
		\end{minipage}%
	\end{center}
	\caption{Video-based animal acts.
		\label{fig_animal_acts}}
\end{figure}
Among the 221 volunteers who participated in the test, 15 participants took the English version and the remainder took the Italian version. Video-based tests offers several advantages with respect to both text-based and image-based tests performed in previous empirical studies on the identification of entanglement in conceptual combinations. In particular, a video-based test reduces language dependence, captures actions more effectively than static images, and leverages audio information to enhance the assessment.

Before analysing the results of the test, it is worth to make some remarks on the choice of its setting. The detection of entanglement in Bell-type cognitive tests, indeed, relies on a precise preparation of both the initial state and the joint measurements. Insufficient initial conditions, as lacking a brief introduction and prior knowledge of the animals and sounds involved, can hinder the ability to identify entanglement. Birds could offer an interesting case study because some species possess unique vocalizations distinct from other individuals.
This individuality makes them well-suited for entanglement experiments involving experts. One could, e.g., consider, two pairs of items of {\it Animal}, namely, ({\it Cuckoo}, {\it Nightingale}) and ({\it Thrush}, {\it Goldfinch}), and two pairs of items of {\it Acts}, ({\it Cuckoos}, {\it Trills}) and ({\it Whistles}, {\it Snails}), without determining a violation of the CHSH inequality in Equation (\ref{chsh}). In other words, cognitive tests on unskilled participants may fail to detect entanglement or yield lower value of the CHSH factor $\Delta_{CHSH}$ in Equation (\ref{chshfactor}). A similar situation is likely to occur if the actions are not typical of the animals. Consider, e.g., two pairs of items of {\it Animal} ({\it Horse}, {\it Bear}) and ({\it Dog}, {\it Camel}), and two pairs of items of {\it Sport} ({\it Racing}, {\it Fishing}) and ({\it Sledding}, {\it Show Jumping}) for the conceptual combination {\it The Animal does Sport}. Also in this case, we expect the CHSH inequality in Equation (\ref{chsh}) not to be violated or to show a little violation. This is why the identification of the items to use in the cognitive test followed strict prescriptions, as follows:

(1) Instance analysis. Identify all possible items of the two concepts under consideration. Search within corpora for pairs of items that appear together, evaluating the meaningfulness of such combinations.

(2) Pair generation. Consider all possible pairs of animals and sounds, verifying that at least one combination is present in the pairs identified in (1).

(3) CAPTCHA creation.
If actions or sounds facilitate the selection, use generative AI to create video-based CAPTCHAs such as Sora AI.
Otherwise, use images and text.

(4) Correlation assessment. Administer a significant number of tests to participants. Calculate the maximum value of $\Delta_{CHSH}$ to establish the correlation between the two concepts.

We are now ready to illustrate the possible judgements of the individuals who participated in the video-based cognitive test.

In the joint measurement $AB$, participants had to choose the best example of {\it The Animal Acts} within the four items: 

($A_1B_1$) {\it The Horse Growls}

($A_2B_2$) {\it The Bear Whinnies}

($A_1B_2$) {\it The Horse Whinnies}

($A_2B_1$) {\it The Bear Growls} 

(sub-test 1 in Figure 1). If the response was $A_1B_1$ or $A_2B_2$, then the measurement $AB$ was attributed the outcome $+1$; if the response was $A_1B_2$ or ($A_2B_1$), then the measurement $AB$ was attributed the outcome $-1$. 

In the joint measurement $AB'$, participants had to choose the best example of {\it The Animal Acts} within the four items: 

($A_1B'_1$) {\it The Horse Snorts}

($A_1B'_2$) {\it The Horse Meows}

($A_2B'_1$) {\it The Bear Snorts}

($A_2B'_2$) {\it The Bear Meows}

(sub-test 2 in Figure 1). If the response was $A_1B'_1$ or $A_2B'_1$, then the measurement $AB'$ was attributed the outcome $+1$; if the response was $A_1B'_2$ or $A_2B'_1$, then the measurement $AB'$ was attributed the outcome $-1$. 

In the joint measurement $A'B$, participants had to choose the best example of {\it The Animal Acts} within the four items:

($A'_1B_1$) {\it The Tiger Growls}

($A'_1B_2$) {\it The Tiger Whinnies}

($A'_2B_1$) {\it The Cat Growls}

($A'_2B_2$) {\it The Cat Whinnies}

(sub-test 3 in Figure 1). If the response was $A'_1B_1$ or $A'_2B_2$, then the measurement $A'B$ was attributed the outcome $+1$; if the response was $A'_1B_2$ or $A'_2B_1$, then the measurement $A'B$ was attributed the outcome $-1$. 

Finally, in the joint measurement $A'B'$, participants had to choose the best example of {\it The Animal Acts} within the four items:

($A'_1B'_1$) {\it The Tiger Snorts}
 
($A'_1B'_2$) {\it The Tiger Meows}

($A'_2B'_1$) {\it The Cat Snorts}

($A'_2B'_2$) {\it The Cat Meows}

(sub-test 4 in Figure 1). If the response was $A'_1B'_1$ or $A'_2B'_2$, then the measurement $A'B'$ was attributed the outcome $+1$; if the response was $A'_1B'_2$ or $A'_2B'_1$, then the measurement $A'B'$ was attributed the outcome $-1$. 

For each joint measurement $XY$, we collected the relative frequencies of the obtained responses which we considered, in the large number limit, as the probability $\mu(X_iY_j)$ that the outcome $X_iY_j=\pm 1$ is obtained in the corresponding measurement, $X=A,A'$, $Y=B,B'$. Table \ref{tab1} reports the judgement probabilities computed in this way. Referring to these probabilities, we can then calculate the expectation values, or correlation functions, of the joint measurements $AB$, $AB'$, $A'B$ and $A'B'$,  as follows:
\begin{eqnarray}
E(A,B)&=&\mu(A_1B_1)-\mu(A_1B_2)-\mu(A_2B_1)+\mu(A_2B_2)=-0.8552\label{AB_Test_1} \\
E(A,B')&=&\mu(A_1B'_1)-\mu(A_1B'_2)-\mu(A_2B'_1)+\mu(A_2B'_2)=0.5204 \label{AB'_Test_1}  \\
E(A',B)&=&\mu(A'_1B_1)-\mu(A'_1B_2)-\mu(A'_2B_1)+\mu(A'_2B_2)=0.7014 \label{A'B_Test_1} \\
E(A',B')&=&\mu(A'_1B'_1)-\mu(A'_1B'_2)-\mu(A'_2B'_1)+\mu(A'_2B'_2)=0.9005 \label{A'B'_Test_1}
\end{eqnarray}
Inserting Equations (\ref{AB_Test_1})--(\ref{A'B'_Test_1}) into Equation (\ref{chsh}), we get
\begin{equation}
\Delta_{CHSH}=E(A',B')+E(A',B)+E(A,B')-E(A,B)=2.9774
\end{equation}
The numerical value $2.9774$ exceeds the classical limit imposed by the CHSH inequality in Equation (\ref{chsh}) and is also above Cirel'son's bound $2\sqrt{2}\approx 2.8284$. A simple check in Table \ref{tab1} reveals that the marginal law conditions in Equations (\ref{marginal1}) and (\ref{marginal2}) is also systematically violated here. This empirical pattern confirms and strengthens the results obtained in previous text-based cognitive tests on the conceptual combination {\it The Animal Acts}, namely \citet{aertssozzo2011,aertsarguellesbeltrangerientesozzo2023b,bertinietal2023}. Table \ref{tab2} reports the judgement probabilities and the CHSH factor in these three text-based tests. 

By comparing the CHSH factors in Tables \ref{tab1} and \ref{tab2}, we can see that a systematic violation of the CHSH inequality occurs in all tests. We also notice that the value of the CHSH factor in the test in \citet{aertssozzo2011} is far from Cirel'son's bound, the test in \citet{aertsarguellesbeltrangerientesozzo2023b} is close to that bound, and the test in \citet{bertinietal2023}, together with the present test violate Cirel'son's bound. That the test in \citet{aertssozzo2011} has a relatively lower deviation from classicality than the other tests might be due to the fact that the introductory text in \citet{aertssozzo2011} encouraged the participants to also take into account emotions and imagination in their judgements. This led them to make choices that are less natural, as \emph{The Bear Meows}, {\it The Cat Growls}, or {\it The Tiger Meows}, thus determining a relatively lower violation of the CHSH inequality in Equation (\ref{chsh}).

As noticed in Sections \ref{intro} and \ref{entanglement}, the empirical results in the video-based cognitive test seem to indicate the presence of entanglement in the combination of {\it Animal} and {\it Acts}. We intend to show in Section \ref{quantum} that this entanglement is present at both the state and the measurement level.
\begin{table}
	\begin{center}
			\begin{tabular}{||c | c ||} 
				\hline
				Participants & 221 \\
				\hline\hline
				\multicolumn{2}{||c||}{Experiment $AB$} \\
				\hline
				{\it Horse Growls} & 0.0452 \\
				\hline
				{\it Horse Whinnies} & 0.8824 \\ 
				\hline
				{\it Bear Growls} & 0.0452 \\ 
				\hline
				{\it Bear Whinnies} & 0.0271 \\ 
				\hline\hline
				\multicolumn{2}{||c||}{Experiment $AB'$} \\
				\hline
				{\it Horse Snorts} & 0.6833 \\ 
				\hline
				{\it Horse Meows} & 0.0226 \\ 
				\hline
				{\it Bear Snorts} & 0.2172 \\ 
				\hline
				{\it Bear Meows} & 0.0770 \\ 
				\hline\hline
				\multicolumn{2}{||c||}{Experiment $A'B$} \\
				\hline
				{\it Tiger Growls} & 0.7919 \\ 
				\hline
				{\it Tiger Whinnies} & 0.0362 \\ 
				\hline
				{\it Cat Growls} & 0.1131 \\ 
				\hline
				{\it Cat Whinnies} & 0.0588 \\ 
				\hline\hline
				\multicolumn{2}{||c||}{Experiment $A'B'$} \\
				\hline
				{\it Tiger Snorts} & 0.0633 \\ 
				\hline
				{\it Tiger Meows} & 0.0452 \\ 
				\hline
				{\it Cat Snorts} & 0.0045 \\ 
				\hline
				{\it Cat Meows} & 0.8869 \\ 
				\hline\hline
				$\Delta_{CHSH}$ & 2.9774 \\ 
				\hline
			\end{tabular}
		\end{center}
	\caption{We report the statistical data collected in the video-based cognitive test presented in Section \ref{test}. The judgement probabilities are hun in substantial agreement with the results obtained in other text-based cognitive tests, namely, \citet{aertssozzo2011}, \citet{aertsarguellesbeltrangerientesozzo2023b} and \citet{bertinietal2023}, presented in Table \ref{tab2}. Also in this case, we get a significant violation of the CHSH inequalitiy in Equation (\ref{chsh}), with a CHSH factor (see Equation (\ref{chshfactor})) exceeding Cirel'son's bound.  \label{tab1}}
\end{table}
\section{Data modelling in Hilbert space\label{quantum}}

In this section, we provide a quantum representation in Hilbert space of the data collected in the video-based cognitive test in Section \ref{test}. We apply the general quantum theoretical framework that we have elaborated for the modelling of any Bell-type test, as we have already done in various articles \citep{aertssozzo2014,aertsetal2019,aertsbeltrangerientesozzo2021,aertsarguellesbeltrangerientesozzo2023b}
.
The quantum theoretical framework for Bell-type situations consists in the implementation of three main steps, as follows.

(i) One identifies in the situation under study, the composite entity, together with the individual entities composing it.

(ii) One recognises in the composite conceptual entity the states, measurements and outcome probabilities that are relevant to the situation under study.

(iii) One represents entities, states, measurements and outcome probabilities using the Hilbert space representation of entities, states, measurements and outcome probabilities of quantum theory. 

(i) The conceptual combination {\it The Animal Acts} is considered as a composite conceptual entity made up of the individual entities {\it Animal} and {\it Acts}. 

(ii) Whenever a person participating in the test reads the introductory text which explains the details of the test and nature of the concepts involved, this set of instructions prepares the composite entity {\it The Animal Acts} in an initial state $p$ which describes the general situation of an animal that produces a recognizable sound. Each participant is then confronted with this uniquely prepared state $p$. More precisely, in the joint measurement $XY$, $X=A,A'$, $Y=B,B'$, each participant interacts with the entity {\it The Animal Acts} in the state $p$ and operates as a measurement context for the entity. This interaction generally changes, in an intrinsically indeterministic way, $p$ into a new state depending on the choice that is made, as a consequence of this `contextual interaction'. E.g., if the participant chooses in $AB$ {\it The Horse Whinnies}, which corresponds to the outcome $A_1B_2$ (see Section \ref{test}), the interaction between the entity  {\it The Animal Acts} in the state $p$ and the (mind of the) participant determines an indeterministic change of state of the entity from $p$ to the state $p_{A_1B_2}$ which describes the more concrete situation of a horse that whinnies. More generally, for every $X=A,A'$, $Y=B,B'$, the joint measurement $XY$ has four possible outcomes $X_iY_j$, where we choose $X_iY_j=+1$ if $i=j$ and $X_iY_j=-1$ if $i \ne j$, and four outcome states, or eigenstates, $p_{X_iY_j}$, describing the state of {\it The Animal Acts} after the outcome $X_iY_j$ occurs in $XY$, $i,j=1,2$. When all responses are collected, a statistics of the outcomes $X_iY_j$ arises which is interpreted, in the large number limit, as the probability $P_{p}(X_iY_j)$ that the outcome $X_iY_j$ is obtained when the joint measurement $XY$ is performed on the composite entity {\it The Animal Acts} in the initial state $p$.

(iii) We have identified the entities, initial state, joint measurements, outcome probabilities and eigenstates that are relevant to the {\it The Animal Acts} situation. Then, we have to work out a quantum representation in Hilbert space to model the data collected in this situation, that is, the entity {\it The Animal Acts} is associated with a complex Hilbert space and the initial state $p$ is represented by a unit vector of this Hilbert space. Next, for every $X=A,A'$, $Y=B,B'$, the joint measurement $XY$ is represented by a self-adjoint operator or, equivalently, by a spectral family, on the Hilbert space whose eigenvectors represent the eigenstates of $XY$, while the outcome probabilities are obtained from Born's rule of quantum probability.

Regarding the Hilbert space representation above, we preliminarily observe that all joint measurements $XY$, $X=A,A'$, $Y=B,B'$, have four outcomes $X_iY_j$, $i,j=1,2$, which entails that the composite entity {\it The Animal Acts} is associated, as an overall entity, with the complex Hilbert space $\mathbb{C}^{4}$ of all ordered 4-tuples of complex numbers. Moreover, each state $p$ of {\it The Animal Acts}  is represented by a unit vector of $\mathbb{C}^{4}$ and each joint measurement on {\it The Animal Acts} is represented by a self-adjoint operator or, equivalently, by a spectral family, on $\mathbb{C}^{4}$. 
On the other hand, for every $i,j=1,2$, each outcome $X_iY_j$ is obtained by juxtaposing the outcomes $X_i$ and $Y_j$, e.g., {\it The Tiger Growls}, is obtained by syntactically juxtaposing the words ``tiger'' and ``growls''. This operation defines a 2-outcome measurement $X$, $X=A,A'$, on the individual entity {\it Animal} and a 2-outcome measurement $Y$, $Y=B,B'$, on the individual entity {\it Acts}. Hence, each of these individual entities is associated with the complex Hilbert space $\mathbb{C}^{2}$ of all ordered pairs of complex numbers. Should we had performed separate measurements on {\it Animal} and {\it Acts}, the Hilbert space formalism would have prescribed that the composite entity {\it The Animal Acts}  would have been associated with the tensor product $\mathbb{C}^{2} \otimes \mathbb{C}^{2}$. But, we remind that we are studying here the identification problem (see Section \ref{entanglement}), that is, the problem of how the composite entity {\it The Animal Acts} can be decomposed into the individual entities {\it Animal} and {\it Acts} in such a way that these individual entities can be recognised from measurements performed on the composite entity. As such, we are doing an operation that is the inverse of what one typically does in Bell-type situations in quantum physics, where one constructs or, better, composes, the measurements on the composite entity from measurements performed on individual entities. 
\begin{table}
	\begin{center}
			\begin{tabular}{||c | c | c | c ||} 
				\hline
				Test & Aerts \& Sozzo (2011) & Aerts et al. (2023) & Bertini et al. (2023) \\ 
				\hline
				Participants & 81 & 81 & 100 \\
				\hline\hline
				\multicolumn{4}{||c||}{Experiment $AB$} \\
				\hline
				{\it Horse Growls} & 0.0494 & 0.0494 & 0.03 \\
				\hline
				{\it Horse Whinnies} & 0.6296 & 0.1235 & 0.91 \\ 
				\hline
				{\it Bear Growls} & 0.0617 & 0.7778 & 0.04\\ 
				\hline
				{\it Bear Whinnies} & 0.2593 & 0.0494 & 0.02 \\ 
				\hline\hline
				\multicolumn{4}{||c||}{Experiment $AB'$} \\
				\hline
				{\it Horse Snorts} & 0.5926 & 0.7160 & 0.83 \\ 
				\hline
				{\it Horse Meows} & 0.0247 & 0.0494 & 0.01 \\ 
				\hline
				{\it Bear Snorts} & 0.2963 & 0.2222 & 0.15 \\ 
				\hline
				{\it Bear Meows} & 0.0864 & 0.0123 & 0.01 \\ 
				\hline\hline
				\multicolumn{4}{||c||}{Experiment $A'B$} \\
				\hline
				{\it Tiger Growls} & 0.7778 & 0.7778 & 0.86 \\ 
				\hline
				{\it Tiger Whinnies} & 0.0864 & 0.0864 & 0 \\ 
				\hline
				{\it Cat Growls} & 0.0864 & 0.0617 & 0.14 \\ 
				\hline
				{\it Cat Whinnies} & 0.0494 & 0.0741 & 0 \\ 
				\hline\hline
				\multicolumn{4}{||c||}{Experiment $A'B'$} \\
				\hline
				{\it Tiger Snorts} & 0.1481 & 0.0864 & 0.01 \\ 
				\hline
				{\it Tiger Meows} & 0.0864 & 0.0617 & 0.02 \\ 
				\hline
				{\it Cat Snorts} & 0.0988 & 0.0247 & 0.02 \\ 
				\hline
				{\it Cat Meows} & 0.6667 & 0.8272 & 0.95 \\ 
				\hline\hline
				$\Delta_{CHSH}$ & 2.4197 & 2.7901 & 3.22 \\ 
				\hline
			\end{tabular}
		\end{center}
	\caption{We report in comparison the statistical data of three text-based cognitive tests, namely, the Aerts \& Sozzo (2011) test \citep{aertssozzo2011} in the first column, the Aerts et al. (2023) test \citep{aertsarguellesbeltrangerientesozzo2023b} in the second column, and the Bertini et al. (2023) test \citep{bertinietal2023} in the third column. The corresponding judgement probabilities are in substantial agreement across the tests and also with the video-based cognitive test in Table \ref{tab1}. All tests exhibit a significant violation of the CHSH inequality.  \label{tab2}}
\end{table}

From a mathematical point of view, the vector spaces $\mathbb{C}^{4}$ and $\mathbb{C}^{2} \otimes \mathbb{C}^{2}$ are isomorphic, where each isomorphism is defined by the relationship between the corresponding orthonormal (ON) bases. The states of {\it The Animal Acts} are represented by unit vectors of $\mathbb{C}^{4}$, hence of $\mathbb{C}^{2} \otimes \mathbb{C}^{2}$, which contains both vectors representing product states and vectors representing entangled states. Moreover, the vector space $L(\mathbb{C}^{4})$ of all linear operators on $\mathbb{C}^{4}$ is isomorphic to the tensor product $L(\mathbb{C}^{2}) \otimes L(\mathbb{C}^{2})$, where $L(\mathbb{C}^{2})$ of all linear operators on $\mathbb{C}^{2}$. Analogously, the tensor product  $L(\mathbb{C}^{2}) \otimes L(\mathbb{C}^{2})$ contains both self-adjoint operators representing product measurements and self-adjoint operators representing entangled measurements (see Section \ref{entanglement}).

Now, let $I: \mathbb{C}^{4} \longrightarrow \mathbb{C}^{2} \otimes \mathbb{C}^{2}$ be an isomorphism mapping a given ON basis of $\mathbb{C}^{4}$ onto a given ON basis of $\mathbb{C}^{2} \otimes \mathbb{C}^{2}$. We say that a state $p$ represented by the unit vector $|p\rangle \in {\mathbb C}^4$ is a `product state with respect to $I$', if two states $p_A$ and $p_B$, represented by the unit vectors $|p_A\rangle \in {\mathbb C}^2$ and $|p_B\rangle \in {\mathbb C}^2$, respectively, exist such that $I|p\rangle=|p_A\rangle\otimes|p_B\rangle$. Otherwise, $p$ is an `entangled state with respect to $I$'. Then, we say that a measurement $e$ represented by the self-adjoint operator ${\mathscr E}$ on ${\mathbb C}^4$ is a `product measurement with respect to $I$', if two measurements $e_X$ and $e_Y$, represented by the self-adjoint operators  ${\mathscr E}_X$ and ${\mathscr E}_Y$, respectively, on ${\mathbb C}^2$ exist such that $I{\mathscr E}I^{-1}={\mathscr E}_X \otimes {\mathscr E}_Y$. Otherwise, $e$ is an `entangled measurement with respect to $I$'. Hence, the notion of entanglement crucially depends on the `isomorphism that is used to identify individual entities within a given composite entity'. 

With reference to a Bell-type setting, one can now prove that, if the joint measurements ${XY}$ and  ${XY'}$, $X=A,A'$, $Y,Y'=B,B'$, $Y \ne Y'$, are product measurements with respect to the isomorphism $I$, then, for every state $p$ of the composed entity, the marginal law condition expressed by Equation (\ref{marginal1}) is satisfied. Analogously, if the joint measurements $XY$ and  $X'Y$, $X,X'=A,A'$, $Y=B,B'$, $X \ne X'$, are product measurements with respect to the isomorphism $I$, then,  for every state $p$ of the composed entity,  the marginal law condition expressed by Equation (\ref{marginal2}) is satisfied. One also proves that, if the marginal law conditions are satisfied in all joint measurements, then a unique isomorphism exists, which can be chosen to be the identity operator \citep{aertssozzo2014}.

It follows from the above that, if the marginal law conditions are violated, then one cannot find a unique isomorphism between $\mathbb{C}^{4}$ and $\mathbb{C}^{2} \otimes \mathbb{C}^{2}$ such that all measurements are product measurements with respect to this isomorphism. In this case, one cannot explain the violation of the CHSH inequality as due to the usual situation in quantum physics where all measurements are product measurements and only the initial pre-measurement state is entangled.\footnote{As anticipated in Section \ref{entanglement}, there are reasons to believe that the marginal law conditions are also violated in typical Bell-type tests on quantum physical entities, which indicates that entangled measurements are involved also in the physical domain. However, the violation of the marginal law conditions in these tests is not large, hence has hardly been reflected about from a theoretical point of view \citep{aertsetal2019}.} Furthermore, if the marginal law conditions are systematically violated, as it occurs in our test (see Table \ref{tab1}), then four distinct isomorphisms $I_{XY}$, exist such that the measurement  ${XY}$ is a product measurement with respect to $I_{XY}$, $X=A,A'$, $Y=B,B'$. As a consequence, there is no unique isomorphism allowing to identify individual entities of a given composite entity. Finally, if we consider a given isomorphism between $\mathbb{C}^{4}$ and $\mathbb{C}^{2} \otimes \mathbb{C}^{2}$ with respect to which identifying individual entities of a composite entity in a given test, then it may happen that both the initial pre-measurement state and all measurements are entangled \citep{aertssozzo2014}. 

Now, the theoretical considerations above allow one to formulate the following hypotheses.

Firstly, the non-classical correlations that violate the CHSH inequality in Equation (\ref{chsh}) in the {\it The Animal Acts} situation can be reasonably attributed to the fact that `the component concepts carry meaning and further meaning is created in the combination process'. Since the violation of the CHSH inequality indicates the presence of entanglement between the individual conceptual entities, then it is reasonable that `it is the quantum structure of entanglement that theoretically captures the meaning that is non-classically created in this case'. This suggests that the initial state $p$ of the composite entity  {\it The Animal Acts} should be an entangled state.

Secondly, in the {\it The Animal Acts} situation, since all joint measurements ${XY}$, $X=A,A'$, $Y=B,B'$, violate the marginal law conditions of Kolmogorovian probability, also these measurements should be entangled measurements. In addition, in each measurement $XY$, all outcomes $X_iY_j$, correspond to combined concepts, though less abstract than {\it The Animal Acts}, e.g., in {\it The Cat Meows}, meaning is created with respect to {\it Cat} and {\it Meows} taken separately, which suggests that all eigenstates $p_{X_iY_j}$, $i,j=1,2$, should also be entangled states.

We are now ready to work out the required quantum representation in Hilbert space of the data in Table \ref{tab1}.

The composite conceptual entity {\it The Animal Acts} is associated with the complex Hilbert space $\mathbb{C}^4$. Let $(1,0,0,0)$, $(0,1,0,0)$, $(0,0,1,0)$ and $(0,0,0,1)\}$ be the unit vectors of the canonical ON basis of $\mathbb{C}^4$, and let us consider the isomorphism $I:\mathbb{C}^4  \longrightarrow \mathbb{C}^2 \otimes \mathbb{C}^2$, where the canonical ON basis of $\mathbb{C}^4$ coincides with the ON basis of the tensor product Hilbert space $\mathbb{C}^2 \otimes \mathbb{C}^2$ made up of the unit vectors $(1,0)\otimes (1,0)$, $(1,0)\otimes (0,1)$, $(0,1)\otimes (1,0)$ and $(0,1)\otimes (0,1)$. 

In the ON bases above, a given state $q$ of the composite entity is represented by the unit vector $|q\rangle=(ae^{i \alpha}, be^{i \beta}, ce^{i \gamma}, de^{i \delta})$, where $a,b,c,d \ge 0$, $a^2+b^2+c^2+d^2=1$, $\alpha$, $\beta$, $\gamma$, $\delta \in \Re$ and $\Re$ is the real line. One easily proves that $|q\rangle$ represents a product state if and only if 
\begin{equation} \label{entanglementcondition}
ade^{i(\alpha+\delta)}-bce^{i(\beta+\gamma)}=0
\end{equation}
Otherwise, $|q\rangle$ represents an entangled state. 

Let us now come to the representation of the initial, or preparation, state $p$ of the composite entity {\it The Animal Acts}. In previous articles, we represented the state $p$ of the conceptual entity {\it The Animal Acts} by the unit vector 
\begin{equation}
|p\rangle=\frac{1}{\sqrt{2}}(0,1,-1,0) \label{singlet}
\end{equation}
which represents the maximally entangled state corresponding to the singlet spin state, as typically done in Bell-type tests in quantum physics (see Section \ref{entanglement}). However, before making this choice here too, it is worth to reflect about the general modelling scheme we adopted to represent the states of conceptual entities (see, e.g., \citet{aertssassolisozzoveloz2021}). In \citet{aertsbeltran2022a} we used the unit vector 
\begin{equation} \label{p_AB}
|p_{AB}\rangle =\sum_{i,j=1,2}\sqrt{\mu(A_iB_j)}|p_{A_iB_j}\rangle
\end{equation}
where $\mu(A_iB_j)$ are the judgement probabilities of the joint measurement $AB$ and the unit vectors $|p_{A_iB_j}\rangle$ form an ON basis of eigenvectors representing the eigenstates of $AB$, $i,j=1,2$, in the 4-dimensional Hilbert space describing the $AB$-measurement situation. This representation is in agreement with the general modelling scheme in \citet{aertssassolisozzoveloz2021}, because the unit vector in Equation (\ref{p_AB}) represents the initial state of {\it The Animal Acts} in a measurement having {\it The Horse Growls}, {\it The Horse Whinnies}, {\it The Bear Growls}, and {\it The Bear Whinnies} as possible outcomes. Analogously, one could use, with obvious changes of symbols, the unit vectors 
\begin{equation} \label{p_AB'}
|p_{AB'}\rangle =\sum_{i,j=1,2}\sqrt{\mu(A_iB'_j)}|p_{A_iB'_j}\rangle
\end{equation}
to represent the initial state of {\it The Animal Acts} when the joint measurement $AB'$ is performed, the unit vector
\begin{equation} \label{p_A'B}
|p_{A'B}\rangle =\sum_{i,j=1,2}\sqrt{\mu(A'_iB_j)}|p_{A'_iB_j}\rangle
\end{equation}
to represent the initial state of {\it The Animal Acts} when the joint measurement $A'B$ is performed, and the unit vector
\begin{equation} \label{p_A'B'}
|p_{A'B'}\rangle =\sum_{i,j=1,2}\sqrt{\mu(A'_iB'_j)}|p_{A'_iB'_j}\rangle
\end{equation}
to represent the initial state of {\it The Animal Acts} when the joint measurement $A'B'$ is performed. Indeed, also the unit vectors in Equations (\ref{p_AB'}), (\ref{p_A'B}) and (\ref{p_A'B'}) reproduce the correct judgement probabilities in the corresponding joint measurements.

How will we then represent the initial state of {\it The Animal Acts} with respect to the overall experiment that tests the CHSH version of Bell's inequalities according to the general modelling scheme in \citet{aertssassolisozzoveloz2021}? The straightforward answer is to take the normalized superposition state represented by the linear combination 
\begin{equation} \label{superposition}
|p_{CHSH}\rangle=\frac{|p_{AB}\rangle+|p_{AB'}\rangle+|p_{A'B}\rangle+|p_{A'B'}\rangle}{\lVert |p_{AB}\rangle+|p_{AB'}\rangle+|p_{A'B}\rangle+|p_{A'B'}\rangle \rVert}
\end{equation} 

Let us recall that the considered basis vectors for each of the joint measurements are not necessarily the same, their relation depending on how these joint measurements relate experimentally to the CHSH form of the Bell's test experiment, and only experiments to test these relations can give us this exact information. This however does not avoid the linear combination in Equation (\ref{superposition}) to be well-defined and representing the initial state of {\it The Animal Acts} with respect to the CHSH form of the Bell's test. 

Let us also recall that the shift from, e.g., an initial, or preparation, state represented by the unit vector $|p_{AB}\rangle$ to an initial, or preparation, state represented by the unit vector $|p_{CHSH}\rangle$ corresponds to (i) an application of a projection operator in Hilbert space which projects onto the one-dimensional subspace generated by the unit vector $|p_{AB}\rangle$, and (ii) a normalization of the resulting projected vector. The opposite shift of the initial, or preparation, state corresponds instead, to a superposition in Hilbert space, as we have explained above.

The considerations above indicate that the unit vector in Equation (\ref{superposition}) is the most appropriate candidate to represent the initial state corresponding to a preparation of the entity {\it The Animal Acts} according to the general modelling scheme in \citet{aertssassolisozzoveloz2021}. However, for the experiments where the data are obtained by calculating relative frequencies of occurrence of different choices, using separately the fours vectors in Equations (\ref{p_AB}), (\ref{p_AB'}), (\ref{p_A'B}) and (\ref{p_A'B'}) comes to the same concerning the predictions made by the model, because relative frequencies of occurrence of outcomes are always determined separately for each of the four joint cases. Or, more concretely, for these experiments we have, e.g., 
\begin{equation}
\mu(X_iY_j) = |\langle p_{X_iY_j}|p_{XY}\rangle|^2 = |\langle p_{X_iY_j}|p_{CHSH}\rangle|^2
\end{equation}
for every $X=A,A'$, $Y=B,B'$, $i,j=1,2$. 

Since the outcomes in cognitive tests with human participants are to a large extent also determined by estimating relative frequencies of occurrence, one can expect that in this case the same, this time almost, equalities. Indeed, since some human participants in the tests will have their choices determined in a totally different way, these equalities do not hold in this case, although even those participants will have the tendency not to let their choice about one of the joint pairs depend on that they gave answers before or after for the other joint pairs, which again makes the effect on outcomes for the two states rather little different, and hence we can speak of `almost equalities’. Hence, given the complexity of the human mind, it is not excluded that the answers given for one of the joint pairs are indeed partly determined by the presence in the same test of the other joint pairs, which means that for psychology experiments we should in principle consider both states giving rise to different predictions for the to be rested probabilities. 

Now, what about assuming the maximally entangled singlet spin state represented by the unit vector $|p\rangle$ in Equation (\ref{singlet}) to be the initial state corresponding to the preparation of {\it The Animal Acts}?

The above is a justifiable choice if we think of a preparation that is very minimal and leads to the test person considering a kind of bare {\it The Animal Acts}. For example, before the instruction sheet is read by the test person, it could have been communicated that it involves measurements involving the sentence ``the animal acts''. For this minimal initial preparation, that maximally entangled state could serve in the quantum theoretical model we are building also according to the general modelling scheme in \citet{aertssassolisozzoveloz2021}. 

We wish to digress briefly in connection with this problem of precisely determining the prepared state, firstly and foremost with the intention of fully clarifying it, and secondly because an element emerges here where our approach shows its strength in connection with a possible use for AI based on quantum structures. Even if we wish to determine the prepared state of an experiment where only the sentence ``the animal acts'' is presented to the test subjects, the singlet spin state is not the best choice to serve as the prepared state. A superposition state in which all animals and all actions that can be performed figure would be the real correct mathematical representation. This would lead however to having to consider a Hilbert space of giant dimension as tensor product of all these possibilities. The human mind is probably not capable of handling this condition, but does make an attempt in that direction. Studies on how concepts are represented in the human mind preferentially point to working with, on the one hand, a precedence for the representation of some leading exemplars (see, e.g., \citet{nosofsky1988} on exemplar theories of concepts) and, on the other hand, with the representation of a prototype of the concept in question (see, e.g., \citet{rosch1973} on prototype theories of concepts). In any case, if these exemplar and prototype theories are correct, they also point to a shortcoming of the human mind in connection with such a representation, and it can be expected that a powerful AI would indeed be able to represent such a giant superposition in every detail. Indeed, in the end, the collection of all animals and all possible actions that each is capable of and that can be conceptually described is still a finite collection. 

Coming back to our quantum representation in Hilbert space of the data presented in Section \ref{test}, there are however also good independent reasons for choosing the unit vector in Equation (\ref{singlet}) to represent the initial state of the entity {\it The Animal Acts}, as we did in previous articles on cognitive entanglement. Indeed, let us recall that our aim is to incorporate as much as possible the entanglement of {\it The Animal Acts} situation into the state preparation. Moreover, the singlet spin state has specific symmetry properties, namely, it is always represented by a unit vector of the form in Equation (\ref{singlet}) independently of the ON basis in which the unit vector is expressed. This would intuitively correspond to the fact that {\it The Animal Acts} expresses a more abstract concept than the corresponding outcomes. Finally, this choice allows one to more easily capture the theoretical connections between entanglement and  meaning, as we will see in the rest of this section.
 
Then, coming to measurements, let us represent the joint measurements ${XY}$, $X=A,A'$, $Y=B,B'$. As we have seen above, each measurement has four outcomes $X_iY_j=\pm 1$ and four eigenstates $p_{X_iY_j}$, $i,j=1,2$. For every $X=A,A'$, $Y=B,B'$, we represent $XY$ by the spectral family defined by the ON basis of the four eigenvectors $|p_{X_iY_j}\rangle$, where we set, for every $i,j=1,2$,
\begin{equation}
|p_{X_iY_j}\rangle=(a_{X_iY_j}e^{i \alpha_{X_iY_j}}, b_{X_iY_j}e^{i \beta_{X_iY_j}}, c_{X_iY_j}e^{i \gamma_{X_iY_j}}, d_{X_iY_j}e^{i \delta_{X_iY_j}}) \label{eigenvectors}
\end{equation}
In Equation (\ref{eigenvectors}), the coefficients are such that $a_{X_iY_j}, b_{X_iY_j}, c_{X_iY_j}, d_{X_iY_j} \ge 0$ and $\alpha_{X_iY_j}, \beta_{X_iY_j},\gamma_{X_iY_j}, \delta_{X_iY_j} \in\Re$. One easily verifies that, for every $X=A,A'$, $Y=B,B'$, $XY$ is a product measurement if and only if all $|p_{X_iY_j}\rangle$s are product vectors. Otherwise, $XY$ is an entangled measurement.

Next, for every $X=A,A'$, $Y=B,B'$, $i,j=1,2$, the probability $P_{p}(X_iY_j)$ of obtaining the outcome $X_iY_j$ in a measurement of ${XY}$ on the composite entity in the state $p$ is given by Born's rule of quantum probability, that is, 
\begin{equation} \label{bornrule}
P_{p}(X_iY_j)=|\langle p_{X_iY_j}|p\rangle|^2
\end{equation}
$i,j=1,2$.

To find a quantum mathematical representation of the data in Table \ref{tab1}, for every measurement $e_{XY}$, the four unit vectors in Equation  (\ref{eigenvectors}) have to satisfy the following three sets of conditions.

(i) Normalization. The eigenvectors in Equation (\ref{eigenvectors})  are unit vectors, that is, for every $X=A,A'$, $Y=B,B'$, $i,j=1,2$,
\begin{equation}
a_{X_iY_j}^2+b_{X_iY_j}^2+c_{X_iY_j}^2+d_{X_iY_j}^2=1
\end{equation}
This corresponds to four conditions for each joint measurement $XY$.

(ii) Orthogonality. The eigenvectors in Equation (\ref{eigenvectors})  are mutually orthogonal, that is,  for every $X=A,A'$, $Y=B,B'$, $i,i',j,j'=1,2$, $i \ne i'$, $j \ne j'$, 
\begin{eqnarray}
\langle p_{X_{i}Y_{j}}|p_{X_{i'}Y_{j'}} \rangle&=&0 \\
\langle p_{X_{i}Y_{j}}|p_{X_{i}Y_{j'}} \rangle&=&0 \\
\langle p_{X_{i}Y_{j}}|p_{X_{i}Y_{j'}} \rangle&=&0
\end{eqnarray}
This corresponds to six additional conditions for each joint measurement $XY$.

 (iii) Probabilities. For every $X=A,A'$, $Y=B,B'$, $i,j=1,2$, the probability $P_{p}(X_iY_j)$ coincides with the empirical probability $\mu(X_iY_j)$ in Table \ref{tab1}, that is,
\begin{equation}
P_p(X_iY_j)=|\langle p_{X_iY_j}|p\rangle|^{2}=\mu(X_iY_j)
\end{equation}
where we have used Born's rule in Equation (\ref{bornrule}). This corresponds to four additional conditions for each joint measurement $XY$.

Finally, let us set, for every $X=A,A'$, $Y=B,B'$, $i,j=1,2$, $\alpha_{X_iY_j}=\beta_{X_iY_j}=\gamma_{X_iY_j}=\delta_{X_iY_j}=\theta_{X_iY_j}$, where $\theta_{X_iY_j}\in \Re$, for the sake of simplicity.

The empirical data in Table \ref{tab1} can be represented in Hilbert space, as follows.

The eigenstates of the measurement ${AB}$ are represented by the unit vectors
\begin{eqnarray}
|p_{A_1B_1}\rangle &=&e^{i 311.20^{\circ}}(0.75,-0.59,-0.29,0) \label{Gsol_HG} \\
|p_{A_1B_2}\rangle &=&e^{i 64.38^{\circ}}(0.01,0.44,-0.88,0.15) \label{Gsol_HW} \\
|p_{A_2B_1}\rangle &=&e^{i 224.23^{\circ}}(0.24,0.30,0,-0.92) \label{Gsol_BG} \\
|p_{A_2B_2}\rangle &=&e^{i 0.53^{\circ}}(0.61,0.60,0.37,0.36) \label{Gsol_BW}
\end{eqnarray}
By applying the entanglement condition in Equation (\ref{entanglementcondition}), we can verify that all eigenstates are entangled, hence ${AB}$ is an entangled measurement. However, one observes that the condition in Equation (\ref{entanglementcondition}) shows a larger deviation from zero in the unit vector $|p_{A_1B_2}\rangle$. We can then say that the eigenstate $p_{A_2B_1}$, corresponding to {\it The Horse Whinnies}, is a `relatively more entangled state', which is reasonably intuitive, as {\it The Horse Whinnies} carries higher meaning with respect to {\it Horse} and {\it Whinnies}. Analogously, the eigenstate $p_{A_2B_2}$, corresponding to {\it The Bear Whinnies}, is a `relatively less entangled state', which is again reasonably intuitive, as {\it The Bear Whinnies} carries lower meaning with respect to {\it Bear} and {\it Whinnies}.

The eigenstates of the measurement ${AB'}$ are represented by the unit vectors
\begin{eqnarray}
|p_{A_1B'_1}\rangle &=&e^{i 0.07^{\circ}}(0.41,0.31,-0.85,0.06)  \label{Gsol_HS} \\
|p_{A_1B'_2}\rangle &=&e^{i 16.59^{\circ}}(-0.01,0.21,0,-0.98) \label{Gsol_HM} \\
|p_{A_2B'_1}\rangle &=&e^{i 131.66^{\circ}}(-0.19,0.92,0.26,0.20) \label{Gsol_BS} \\
|p_{A_2B'_2}\rangle &=&e^{i 180.20^{\circ}}(0.89,0.05,0.45,0) \label{Gsol_BM}
\end{eqnarray}
Also in this case, all eigenstates are entangled, hence ${AB'}$ is an entangled measurement. However, the condition in Equation (\ref{entanglementcondition}) shows a larger deviation from zero in the unit vector $|p_{A_1B'_1}\rangle$. We can then say that the eigenstate $p_{A_1B'_1}$, corresponding to {\it The Horse Snorts}, is a `relatively more entangled state', which is reasonably intuitive, as {\it The Horse Snorts} carries higher meaning with respect to {\it Horse} and {\it Snorts}. Analogously, the eigenstates $p_{A_2B'_1}$ and $p_{A_2B'_2}$, corresponding to {\it The Horse Meows} and {\it The Bear Meows}, are `relatively less entangled states', which is again reasonably intuitive, as {\it The Horse Meows} and {\it The Bear Meows} carry lower meaning with respect to {\it Horse} and {\it Meows} and {\it Bear} and {\it Meows}, respectively.

The eigenstates of the measurement ${A'B}$ are represented by the unit vectors
\begin{eqnarray}
|p_{A'_1B_1}\rangle &=&e^{i 32.52^{\circ}}(0.20,0.35,-0.90,0.14)\label{Gsol_TG} \\
|p_{A'_1B_2}\rangle &=&e^{i 174.92^{\circ}}(0.98,-0.09,0.18,0) \label{Gsol_TW} \\
|p_{A'_2B_1}\rangle &=&e^{i 0.39^{\circ}}(0.01,0.86,0.39,0.32) \label{Gsol_CG} \\
|p_{A'_2B_2}\rangle &=&e^{i 205.95^{\circ}}(0.03,0.35,0,-0.94) \label{Gsol_CW}
\end{eqnarray}
All eigenstates are entangled, hence ${A'B}$ is an entangled measurement. However, the condition in Equation (\ref{entanglementcondition}) shows a larger deviation from zero in the unit vector $|p_{A'_1B_1}\rangle$. We can then say that the eigenstate $p_{A'_1B_1}$, corresponding to {\it The Tiger Snorts}, is a `relatively more entangled state', which is reasonably intuitive, as {\it The Tiger Snorts} carries higher meaning with respect to {\it Tiger} and {\it Snorts}. Analogously, the eigenstates $p_{A'_1B_2}$ and $p_{A'_2B_2}$, corresponding to {\it The Tiger Whinnies} and {\it The Cat Whinnies}, are `relatively less entangled states', which is again reasonably intuitive, as {\it The Tiger Whinnies} and {\it The Cat Whinnies} carry lower meaning with respect to {\it Tiger} and {\it Whinnies} and {\it Cat} and {\it Whinnies}, respectively.

Finally, the eigenstates of the measurement ${A'B'}$ are represented by the unit vectors
\begin{eqnarray}
|p_{A'_1B'_1}\rangle &=&e^{i 99.21^{\circ}}(0.73,-0.63,-0.27,0) \label{Gsol_TS} \\
|p_{A'_1B'_2}\rangle &=&e^{i 20.16^{\circ}}(0.27,0.31,0.01,-0.91)\label{Gsol_TM} \\
|p_{A'_2B'_1}\rangle &=&e^{i 0.69^{\circ}}(0.62,0.53,0.44,0.38) \label{Gsol_CS} \\
|p_{A'_2B'_2}\rangle &=&e^{i 353.58^{\circ}}(0.09,0.47,-0.86,0.18) \label{Gsol_CM}
\end{eqnarray}
Also in this case, all eigenstates are entangled, hence ${A'B'}$ is an entangled measurement. However, the condition in Equation (\ref{entanglementcondition}) shows a larger deviation from zero in the unit vector $|p_{A'_1B'_1}\rangle$. We can then say that the eigenstate $p_{A'_2B'_2}$, corresponding to {\it The Cat Meows}, is a `relatively more entangled state', which is reasonably intuitive, as {\it The Cat Meows} carries higher meaning with respect to {\it Cat} and {\it Meows}. Analogously, the eigenstate $p_{A'_2B'_1}$, corresponding to {\it The Cat Snorts}, is a `relatively less entangled state', which is again reasonably intuitive, as {\it The Cat Snorts} carries lower meaning with respect to {\it Cat} and {\it Snorts}.

We have thus completed the quantum mathematical representation of the data on the video-based cognitive test in Section \ref{test}. This representation, however, also suggests relevant considerations. This will be the content of Section \ref{explanation}.

\section{Entanglement as a mechanism of contextual updating\label{explanation}}
The quantum theoretical modelling elaborated in Section \ref{quantum} allows one to draw some interesting conclusions regarding the appearance of entanglement in the combination of natural concepts and, more important, on the nature of this entanglement. We stress, however, that these conclusions are independent of the domain, cognitive or physical, where entanglement is applied, as we have extensively explained in Sections \ref{entanglement} and \ref{marginal}. This means that the results obtained in the present article may also shed new light on the nature of physics entanglement.

Firstly, we have explicitly worked out a quantum mathematical model in Hilbert space which explains the violation of the CHSH inequality in the video-based cognitive test on {\it The Animal Acts} as a demonstration of the presence of quantum entanglement. More explicitly, the individual conceptual entities {\it Animal} and {\it Acts} entangle when they combine to form the combined, or composite, conceptual entity {\it The Animal Acts}. The reason of this entanglement is that both concepts {\it Animal} and {\it Acts} carry meaning. But, also the combination {\it The Animal Acts} carries its own meaning. And, the meaning of {\it The Animal Acts} is not attributed by separately attributing meaning to {\it Animal} and {\it Acts}, as would be prescribed by a classical compositional semantics.

Secondly, we have seen that, not only the initial state of the composite entity is entangled, but also all joint measurements are entangled, in the quantum representation of {\it The Animal Acts} situation. This result is due to the violation of the marginal law conditions of Kolmogorovian probability in the cognitive test which forbids concentrating all the entanglement of the state-measurement situation into the state of the composite entity, as we have seen in Section \ref{entanglement}, where we have also provided arguments to conclude that no-signaling is at place. 

Thirdly, in each joint measurement, all eigenstates are entangled. This result is due to the fact that, in each joint measurement, all possible outcomes correspond to combinations of concepts, e.g., {\it The Bear Snorts} is itself a combination of the concepts {\it Bear} and {\it Snorts}, hence it is reasonable to expect that a non-classical mechanism of meaning attribution, similar to {\it The Animal Acts}, occurs.

Fourthly, in each joint measurement, some eigenstates, which are the final states at the end of the joint measurement, exhibit a relatively higher degree of entanglement than others, which can exactly be explained by the fact that entanglement captures meaning attribution, hence higher degrees of entanglement correspond to higher meaning attribution, thus higher judgement probabilities. For example, the eigenstates corresponding to {\it The Horse Whinnies}, {\it The Horse Snorts}, {\it The Tiger Growls} and {\it The Cat Meows} are the states that exhibit the highest degree of entanglement in the corresponding joint measurement. As observed in Section \ref{quantum}, this can be explained as due to the fact that higher meaning is attributed in the combination process for these items. As a matter of fact, these items score the highest probability of   
being judged as a good example of the conceptual combination {\it The Animal Acts}. By contrast, the eigenstates corresponding to {\it The Bear Whinnies}, {\it The Horse Meows}, {\it The Tiger Whinnies} and {\it The Cat Snorts} are the states that exhibit the lowest degree of entanglement, close to product states, in the corresponding joint measurement. Again this can be explained as due to the fact that lower meaning is attributed in the combination process for these items. As a matter of fact, these items score the lowest probability of being judged as a good example of the conceptual combination {\it The Animal Acts}.

Fifthly, by looking at the CHSH factor in Table \ref{tab1}, we notice that the video-based cognitive test violates the CHSH inequality by an amount that exceeds the value $2\sqrt{2}\approx 2.8284$, i.e. Cirel'son's bound, which is usually believed to be the theoretical limit to represent in Hilbert space the statistical correlations that are observed in Bell-type tests by pushing all entanglement into the state of the composite entity and considering only product measurements. Again, this is due to the fact that the joint measurements are actually entangled, rather than product, measurements. As we have illustrated in Sections \ref{entanglement} and \ref{marginal}, independently of the physical or cognitive domain of reference, if one allows entangled measurements, then a Hilbert space representation is possible also for Bell-type tests which violate Cirel'son's bound \citep{aertsarguellesbeltrangerientesozzo2023b}.

We would like to conclude the present article by deepening the mechanism of meaning attribution to concepts and its relationship with quantum entanglement. As we have seen above, the appearance of entanglement in {\it The Animal Acts} is due to the fact that people
attribute meaning to the combination {\it The Animal Acts} as a whole entity, without firstly attributing meaning to {\it Animal} and {\it Acts} and then combining these separate meanings into a meaning for {\it The Animal Acts}.  One way to characterize this process of meaning assignment is the following. A concept carries a meaning, and a second concept carries a meaning, however, the combination of these two concepts also carries its proper meaning, and this is not the simple combination of the two meanings of the component concepts as prescribed by a classical compositional semantics. On the contrary, `the new emergent meaning of the combined concept arises in a complex contextual way', in which the whole of the context relevant to the combination plays a role.

The above is even more evident if one considers an entire text produced by human language and its meaning relationship with the words (concepts) composing it. Each time a word (concept) is added to a text, one can speak of an `updating of contextuality', and this updating continues to occur until the end of the text that contains all the words. As we have argued in \citet{aertsarguellesbeltrangerientesozzo2023b,aertsarguellesbeltrangerientesozzo2023a}, this mechanism of `contextual updating' to attribute meaning has to be carried by an entangled state, because this is exactly how entangled states are formed in the tensor product of Hilbert spaces. In other words, it is these entangled states that accomplish the contextual updating in the mathematical formalism of quantum theory. The deep structural similarities between physical and cognitive domains, suggest that the mechanism of contextual updating could also explain, better than the `spooky action at a distance' mechanism, how entanglement is produced in physics \citep{aertsarguellesbeltrangerientesozzo2023a}. 

Coming back to {\it The Animal Acts} situation, the concept {\it Animal} is an abstraction of all possible animals and the concept {\it Acts} is an abstraction of all possible sounds produced by animals. But, people do not construct the meaning of {\it The Animal Acts} by separately considering abstractions of animals and abstractions of acts and then combining these abstractions. On the contrary, they take directly abstractions of animals making a sound, and this occurs in a coherent way that is represented by a superposed, more precisely, entangled, state. This is exactly what we have defined above as the mechanism of contextual updating.

\section*{Acknowledgements}
This work was supported by the project ``New Methodologies for Information Access and Retrieval with Applications to the Digital Humanities'', scientist in charge S. Sozzo, financed within the fund ``DIUM -- Department of Excellence 2023--27'' and by the funds that remained at the Vrije Universiteit Brussel at the completion of the ``QUARTZ (Quantum Information Access and Retrieval Theory)'' project, part of the ``European Union Marie Sklodowska-Curie Innovative Training Network 721321'', with Diederik Aerts as principle investigator for the Brussels part of the network.




\begin{thebibliography}{99}
\setlength{\itemsep}{-0.28 mm}

\bibitem[Accardi \& Fedullo, 1982]{accardifedullo1982} Accardi, L. and Fedullo, A. (1982). On the statistical meaning of the complex numbers in quantum mechanics. {\it Nuovo Cimento 34}, 161--172.

\bibitem[Adenier \& Khrennikov, 2007]{adenierkhrennikov2007} Adenier, G. and Khrennikov, A. Y. (2007). Is the fair sampling assumption supported by EPR experiments? {\it Journal of Physics B: Atomic, Molecular and Optical Physics 40}, 131--141.

\bibitem[Adenier \& Khrennikov, 2017]{adenierkhrennikov2016} Adenier, G. and Khrennikov, A. Y. (2017). Test of the no-signaling principle in the Hensen loophole-free CHSH experiment. {\it Fortschritte der Physik (Progress in Physics) 65}, 1600096.

\bibitem[Aerts, 1982]{aerts1982} Aerts, D. (1982). Example of a macroscopical classical situation that violates Bell inequalities. {\it Lettere al Nuovo Cimento 34}, 107--111.

\bibitem[Aerts, 1983]{aerts1983} Aerts, D. (1983). The description of one and many physical systems. In Gruber, C. (Ed), {\it Foundations of Quantum Mechanics}, pp. 63--148. Lausanne: Association Vaudoise des Chercheurs en Physique. 

\bibitem[Aerts, 1991]{aerts1991} Aerts, D. (1991). Aerts, D. A mechanistic classical laboratory situation violating the Bell inequalities with $2\sqrt{2}$, exactly `in the same way' as its violations by the EPR experiments. {\it Helvetica Physica Acta 64}, 1--23.

\bibitem[Aerts, 2009a]{aerts2009a} Aerts, D. (2009a). Quantum structure in cognition. {\it Journal of Mathematical Psychology 53}, 314--348.


\bibitem[Aerts, 2009b]{aerts2009b} Aerts, D. (2009b).  Quantum particles as conceptual entities: A possible explanatory framework for quantum theory. {\it Foundations of Science 14}, 361--411.


\bibitem[Aerts, 2014]{aerts2014} Aerts, D. (2014). Quantum and concept combination, entangled Measurements, and prototype theory. {\it Topics in Cognitive Science 6},   129--137.

\bibitem[Aerts et al., 2018]{aertsetal2018} Aerts, D., Aerts Argu\"{e}lles, J., Beltran, L., Beltran, L., Distrito, I.,  Sassoli de Bianchi, M., Sozzo, S. and Veloz, T. (2018). Towards a quantum World Wide Web. {\it Theoretical Computer Science 752}, 116--131. 

\bibitem[Aerts et al., 2017]{aertsarguellesbeltransassolisozzoveloz2017} Aerts, D., Argu\"{e}lles, J., Beltran, L., Beltran, L., Sassoli de Bianchi, M., Sozzo, S. and Veloz, T. (2017). Testing quantum models of conjunction fallacy on the World Wide Web, {\it International Journal of Theoretical Physics 56}, 3744--3756.

\bibitem[Aerts et al., 2019]{aertsetal2019} Aerts, D., Aerts Argu\"{e}lles, J., Beltran, L., Geriente, S., Sassoli de Bianchi, M., Sozzo, S. and Veloz, T. (2019). Quantum entanglement in physical and cognitive systems: A conceptual analysis and a general representation. {\it The European Physical Journal Plus 134}, 493.

\bibitem[Aerts et al., 2023a]{aertsarguellesbeltrangerientesozzo2023a} Aerts, D., Aerts Argu\"{e}lles, J., Beltran, L., Geriente, S., Sozzo, S. (2023a). Entanglement as a method to reduce uncertainty. {\it International Journal of Theoretical Physics 62}, 145.

\bibitem[Aerts et al., 2023b]{aertsarguellesbeltrangerientesozzo2023b} Aerts, D., Aerts Argu\"elles, J., Beltran, L., Geriente, S., Sozzo, S. (2023b)
Entanglement in cognition. Violating Bell inequalities beyond Cirel'son's bound. In Plotnitsky, A., Haven, E. {\it The Quantum-Like Revolution}, pp. 299--326. Cham: Springer.


\bibitem[Aerts \& Beltran, 2022]{aertsbeltran2022a} Aerts, D. and Beltran, L. (2022). Are words the quantua of human language? Extending the domain of quantum cognition. {\it Entropy 24}, 6.

\bibitem[Aerts et al., 2021] {aertsbeltrangerientesozzo2021} Aerts, D., Beltran, L., Geriente, S. and Sozzo, S. (2021). Quantum-theoretic Modeling in Computer Science. A Complex Hilbert Space Model for Entangled Concepts in Corpuses of Documents. {\it International Journal of Theoretical Physics 60}, 710--726.

\bibitem[Aerts et al., 2013]{aertsbroekaertgaborasozzo2013} Aerts, D., Broekaert, J., Gabora, L. and Sozzo, S. (2013). Quantum structure and human thought. {\it Behavioral and Brain Sciences 36}, 274--276. 

\bibitem[Aerts Haven \& Sozzo, 2018]{haven2018} Aerts, D., Haven, E. and Sozzo, S. (2018). A proposal to extend expected utility in a quantum probabilistic framework, {\it Economic Theory 65}, 1079--1109.


\bibitem[Aerts Gabora \& Sozzo, 2013]{aertsgaborasozzo2013} Aerts, D., Gabora, L. and Sozzo, S. (2013). Concepts and their dynamics: a quantum theoretic modeling of human thought. {\it Topics in Cognitive Science 5}, 737--772.

\bibitem[Aerts \& Sassoli de Bianchi, 2021]{aertssassolidb2021} Aerts, D. and Sassoli de Bianchi, M. (2021). Violation of the Bell-CHSH inequality and marginal laws in a single-entity Bell-test experiment. {\it Journal of Mathematical Physics 62}, 092103.


\bibitem[Aerts Sassoli de Bianchi \& Sozzo, 2016]{aertssassolisozzo2016} Aerts, D., Sassoli de Bianchi, M. and Sozzo, S. (2016). On the foundations of the Brussels operational-realistic approach to cognition. {\it Frontiers in Physics 4}, 17.


\bibitem[Aerts et al., 2021]{aertssassolisozzoveloz2021} Aerts, D., Sassoli de Bianchi, M., Sozzo, S. and Veloz, T. (2021). Modeling human decision-making: An overview of the Brussels quantum approach. {\it Foundations of Science 26}, 27--54.

\bibitem[Aerts \& Sozzo, 2011]{aertssozzo2011} Aerts, D. and Sozzo, S. (2011). Quantum structure in cognition: Why and how concepts are entangled. {\it Quantum Interaction. Lecture Notes in Computer Science 7052}, 116--127. 

\bibitem[Aerts \& Sozzo, 2014]{aertssozzo2014} Aerts, D. and Sozzo, S. (2014). Quantum entanglement in concept combinations. {\it International Journal of Theoretical Physics 53}, 3587--3603.

\bibitem[Aerts Argu\"elles, 2018]{arguelles2018} Aerts Argu\"{e}lles, J. (2018). The heart of an image: Quantum superposition and entanglement in visual perception. {\it Foundations of Science 23}, 757--778.

\bibitem[Aerts Argu\"elles \& Sozzo, 2020]{arguellessozzo2020} Aerts Argu\"{e}lles, J. and Sozzo, S. (2020). How images combine meaning: Quantum entanglement in visual perception. {\it Soft Computing 24}, 10277--10286.

\bibitem[Aerts, 2005]{aerts2005} Aerts, S. (2005). A realistic device that simulates the non-local PR box without communication. {\it ArXiv quant-ph/0504171}.


\bibitem[Bedhorz, 2017]{bednorz2017} Bednorz A. (2017). Analysis of assumptions of recent tests of local realism. {\it Physical Review A 95}, 042118.


\bibitem[Bell, 1964]{bell1964} Bell, J. (1964). On the Einstein Podolsky Rosen paradox. {\it Physics 1}, 195--200.

\bibitem[Beltran \& Geriente, 2019]{beltrangeriente2019} Beltran, L. and Geriente, S. (2019). Quantum entanglement in corpuses of documents. {\it Foundations of Science 24}, 227--246.

\bibitem[Bertini Lemporini \& Moriani, 2023]{bertinietal2023} Bertini, C., Leporini, R., Moriani, S. (2023) Quantum entanglement and encoding algorithm. {\it International Journal of Theoretical Physics 62}, 149, 1--10.

\bibitem[Bohm, 1951]{bohm1951} Bohm, D. (1951). \emph{Quantum Theory}. New-York: Prentice-Hall.

\bibitem[Brunner et al., 2014]{brunner2014} Brunner, N., Cavalcanti, D., Pironio, S., Scarani, V. and Wehner, S. (2014). Bell nonlocality. {\it Reviews of Modern Physics 86}, 419 and 839.

\bibitem[Bruza et al., 2009a]{bruza2009} Bruza, P., Kitto, K., Nelson, D. and McEvoy, C. (2009a). Is there something quantum-like about the human mental lexicon? {\it Journal of Mathematical Psychology 53}, 362--377.

\bibitem[Bruza et al., 2009b]{bruzakittonelsonmcevoy2009} Bruza, P., Kitto, K., Nelson, D. and McEvoy, C. (2009b). Extracting spooky-activation-at-a-distance from considerations of entanglement. In Bruza, P., Sofge, D., Lawless, W., van Rijsbergen, K., Klusch, M. (Eds), {\it Quantum Interaction. QI 2009. Lecture Notes in Computer Science 5494}, pp. 71--83. Berlin: Springer.


\bibitem[Bruza et al., 2016]{bruzakittorammsitbon2015} Bruza, P., Kitto, K., Ramm, B. and Sitbon, L. (2015). A probabilistic framework for analysing the compositionality of conceptual combinations. {\it Journal of Mathematical Psychology 67}, 26--38. 

\bibitem[Buccio Melucci \& Song, 2011]{bucciomeluccisong2011} Di Buccio, E., Melucci, M. and Song, D. (2011). Towards predicting relevance using a quantum-like framework.  In P. Clough et al. (Eds.), {\it Advances in Information Retrieval. ECIR 2011. Lecture Notes in Computer Science 6611}, pp. 755--758. Berlin: Springer.

\bibitem[Busemeyer \& Bruza, 2012]{busemeyerbruza2012} Busemeyer, J. and Bruza, P. (2012). {\it Quantum Models of Cognition and Decision}. Cambridge: Cambridge University Press.

\bibitem[Carnap, 1947]{carnap1947} Carnap, R. (1947). {\it Meaning and Necessity: A Study in Semantics and Modal Logic}. Chicago: The University of Chicago Press.

\bibitem[Christensen et al., 2013]{urbana2013} Christensen, B. G., McCusker, K. T., Altepeter, J., Calkins, B., Gerrits, T., Lita, A., Miller, A., Shalm, L. K., Zhang, Y., Nam, S. W., Brunner, N., Lim, C. C. W., Gisin, N. and Kwiat, P. G. (2013). Detection-loophole-free test of quantum nonlocality, and applications. {\it Physical Review Letters 111}, 1304--1306.


\bibitem[Cirelson, 1980]{cirelson1980} Cirel'son, B. S. (1980). Quantum generalizations of Bell's Inequality. {\it Letters in Mathematical Physics 4}, 93--100. 

\bibitem[Cirelson, 1993]{cirelson1993} Cirel'son, B. S. (1993). Some results and problems on quantum Bell-type inequalities. {\it Hadronic Journal Supplement 8}, 329--345.


\bibitem[Clauser et al., 1969]{chsh1969} Clauser, J. F., Horne, M. A., Shimony, A. and Holt, R.A. (1969). Proposed experiment to test local hidden-variable theories. {\it Physical Review Letters 23}, 880--884.

\bibitem[Coecke Sadrzadeh \& Clark, 2010]{coecke2010} Coecke, B., Sadrzadeh, M. and Clark, S. (2010). Mathematical foundations for a compositional distributional model of meaning. {\it Linguistic Analysis 36}, 345--384.

\bibitem[Dalla Chiara et al., 2015a]{dallachiaragiuntininegri2015b} Dalla Chiara, M. L., Giuntini, R., Leporini, R., Negri, E. and Sergioli, G. (2015a). Quantum information, cognition, and music. {\it Frontiers in Psychology 6},  1583. 

\bibitem[Dalla Chiara et al., 2015b]{dallachiaraetal2015} Dalla Chiara, M.L., Giuntini, R., Leporini, R., Sergioli, G. (2015b). A first-order epistemic quantum computational semantics with relativistic-like epistemic effects. {\it Fuzzy Sets and Systems 298}, 69--90.

\bibitem[Dalla Chiara et al., 2015c]{dallachiaragiuntininegri2015a} Dalla Chiara, M. L., Giuntini, R. and Negri, E. (2015c). A quantum approach to vagueness and to the semantics of music. {\it International Journal of Theoretical Physics 54}, 4546--4556. 

\bibitem[de Barros, 2015]{barros2015} de Barros, J. A., Dzhafarov, E. N., Kujala, J. V. and Oas, G. (2015). Measuring observable quantum contextuality. {\it Quantum Interaction. Lecture Notes in Computer Science 9535}, 36--47. 

\bibitem[De Raedt Michielsen \& Jin, 2012]{deraedt2012} De Raedt, H., Michielsen, K. and Jin, F. (2012). Einstein-Podolsky-Rosen-Bohm laboratory experiments: Data analysis and simulation. {\it AIP Conference Proceedings 1424}, 55--66.

\bibitem[Dzhafarov \& Kujala, 2016]{dzhafarov2016} Dzhafarov, E. N. and Kujala, J. V. (2016). Context-content systems of random variables: The Contextuality-by-Default theory. {\it
Journal of Mathematical Psychology 74}, 11--33.

\bibitem[Einstein Podolsky \& Rosen, 1935]{epr1935} Einstein, A., Podolsky, B. and Rosen, N. (1935). Can Quantum-Mechanical Description of Physical Reality Be Considered Complete? {\it Physical Review 47}, 777--780.

\bibitem[Frommholz et al., 2010]{frommholzetal2010} Frommholz, I., Larsen, B., Piwowarski, B., Lalmas, M. and Ingwersen, P. (2010). Supporting polyrepresentation in a quantum-inspired geometrical retrieval framework. {\it IIiX '10 Proceedings of the Third Symposium on Information Interaction in Context}, pp. 115--124. 


\bibitem[Genovese, 2005]{genovese2005} Genovese, M. (2005). Research on hidden variable theories. A review of recent progresses. {\it Physics Reports 413}, 319--396.

\bibitem[Giustina et al., 2013]{vienna2013} Giustina, M., Mech, A., Ramelow, S., Wittmann, B., Kofler, J., Beyer, J., Lita, A., Calkins, B., Gerrits, T., Woo Nam, S., Ursin, R. and Zeilinger, A. (2013). Bell violation using entangled photons without the fair-sampling assumption. {\it Nature 497}, 227--230. 

\bibitem[Gronchi \& Strambini, 2017]{gronchistrambini2017} Gronchi, G. and Strambini, E. (2017). Quantum cognition and Bell's inequality: A model for probabilistic judgment bias. {\it Journal of Mathematical Psychology 78}, 65--75. 

\bibitem[Haven \& Khrennikov, 2013]{havenkhrennikov2013} Haven, E. \& Khrennikov, A.Y. (2013). \emph{Quantum Social Science}, Cambridge: Cambridge University Press.

\bibitem[Horodecki Horodecki \& Horodecki, 2009]{horodecki2009} Horodecki, R., Horodecki, P., Horodecki, M. and Horodecki, K.(2009). Quantum entanglement. {\it Review of Modern Physics 81}, 865.

\bibitem[Khrennikov, 2010]{khrennikov2010} Khrennikov, A. (2010). {\it Ubiquitous Quantum Structure}. Berlin: Springer.

\bibitem[Kupczynski, 2017]{kupczynski2017} Kupczynski, M. (2017). Is Einsteinian no-signalling violated in Bell tests? {\it Open Physics 15}, 739--753.

\bibitem[Kvam et al., 2015]{kwampleskacbusemeyer2015} Kvam, P., Pleskac, T., Yu, S. and Busemeyer, J. (2015). Interference effects of choice on confidence. {\it Proceedings of the National Academy of Science of the USA 112}, 10645--10650. 

\bibitem[Melucci, 2015]{melucci2015} Melucci, M. (2015). {\it Introduction to Information Retrieval and Quantum Mechanics}. Berlin Heidelberg: Springer.

\bibitem[Nosofski, 1988]{nosofsky1988} Nosofsky, R. (1988). Exemplar-based accounts of relations between classification, recognition, and typicality. {\it Journal of Experimental Psychology: Learning, Memory, and Cognition 14}, 700--708.

\bibitem[Peres \& Terno, 2004]{peresterno2004} Peres A. and Terno D. (2004). Quantum information and relativity theory. {\it Review of Modern Physics 76}, 93--123. 

\bibitem[Pisano \& Sozzo, 2020]{pisanosozzo2020}  Pisano, R. and Sozzo, S. (2020). Unified theory of human judgements and decision-making under uncertainty. {\it Entropy 22}, 738.

\bibitem[Pitowsky, 1989]{pitowsky1989} Pitowsky, I. (1989). {\it Quantum Probability, Quantum Logic}, Lecture Notes in Physics 321. Berlin: Springer.

\bibitem[Piwowarski et al., 2010]{piwowarskietal2010} Piwowarski, B., Frommholz, I., Lalmas, M. and van Rijsbergen, K. (2010). What can quantum theory bring to information retrieval. {\it Proceedings of the 19th ACM international conference on Information and knowledge management}, pp. 59--68.

\bibitem[Pothos \& Busemeyer, 2009]{pothosbusemeyer2009} Pothos, E. and Busemeyer, J. (2009). A quantum probability explanation for violations of `rational' decision theory. {\it Proceedings of the Royal Society B 276}, 2171--2178. 

\bibitem[Rosch, 1973]{rosch1973} Rosch, E. H. (1973). Natural categories. {\it Cognitive Psychology 4}, 328--350.

\bibitem[Schr\"odinger, 1935]{schrodinger1935} Schr\"odinger, E. (1935). Discussion of probability relations between separated systems.{\it Mathematical Proceedings of the Cambridge Philosophical Society 31}, 555--563.

\bibitem[Zellh{\" o}fer et al., 2011]{ingo2011}  Zellh{\" o}fer, D., Frommholz, I., Schmitt, I., Lalmas, M. and van Rijsbergen, K. (2011). Towards quantum-based DB+IR processing based on the principle of polyrepresentation. In P. Clough et al. (Eds.), {\it Advances in Information Retrieval, ECIR 2011, LNCS 6611}, pp. 729--732. Berlin: Springer.

\end{thebibliography}


\end{document}